\documentclass{aa}  
\usepackage{natbib}
\usepackage{multirow}
\bibpunct{(}{)}{;}{a}{}{,}
\usepackage{graphicx}
\usepackage{color}
\usepackage{txfonts}
\begin{document} 

\title{HR 10: A main-sequence binary with circumstellar
envelopes around both components }
\subtitle{Discovery and analysis\thanks{Partially based on 
observations obtained with PIONIER/VLT (ESO, Paranal, Chile), FIES/NOT, HERMES/Mercator, 
HARPS-N/TNG and UES/WHT (La Palma, Spain), FEROS/2.2-m ESO-MPIA (La Silla, Chile),
CS21/Harlan J. Smith Telescope (McDonald Observatory, US) and UHRF/3.6-m AAT
(Anglo Australian Observatory), and archival data from HARPS/3.6-m ESO and 
UVES/VLT (ESO archive), and HIRES/Keck 1 (Keck archive).}}


   \author{B. Montesinos
          \inst{1}
          \and         
          C. Eiroa          
          \inst{2,3}
          \and
          J. Lillo-Box              
          \inst{4,1}
          \and 
          I. Rebollido
          \inst{2}
          \and
          A.A. Djupvik
          \inst{5}
          \and
          O. Absil
          \inst{6}\fnmsep\thanks{F.R.S.-FNRS Research Associate} 
          \and
          S. Ertel
          \inst{7,8}
          \and
          L. Marion
          \inst{6}
          \and
          J.J.E. Kajava
          \inst{9}
          \and
          S. Redfield
          \inst{10}
          \and
          H. Isaacson
          \inst{11}
          \and
          H. C\'anovas
          \inst{12}
          \and
          G. Meeus
          \inst{2}
          \and
          I. Mendigut\'{\i}a
          \inst{1}
          \and
          A. Mora
          \inst{13}
          \and
          P. Rivi\`ere-Marichalar
          \inst{14}
          \and
          E. Villaver
          \inst{2}
          \and
          J. Maldonado
          \inst{15}
          \and
          T. Henning
          \inst{16}
   }
            \institute{Departmento de Astrof\'{\i}sica, Centro de Astrobiolog\'{\i}a
                      (CAB, CSIC-INTA), ESAC Campus, Camino Bajo del Castillo s/n,
                       28692 Villanueva de la Ca\~nada, Madrid, Spain\\
                       \email{bmm@cab.inta-csic.es}
                       \and
             Universidad Aut\'onoma de Madrid, Dpto. F\'{\i}sica Te\'orica, Facultad 
             de Ciencias, Campus de Cantoblanco, 28049 Madrid, Spain
                       \and
             Observatorio Astr\'onomico de Calar Alto, CAHA, 04550 G\'ergal, Almer\'{\i}a, Spain
                       \and
             European Southern Observatory (ESO), Alonso de C\'ordova 3107, Vitacura, Casilla 19001, Santiago de Chile, Chile
                       \and
             Nordic Optical Telescope, Apartado 474, 38700 Santa Cruz de La Palma, Santa Cruz de Tenerife, Spain
                       \and
             Space sciences, Technologies, and Astrophysics Research (STAR) Institute, University of Li\`ege, Belgium
                       \and
             Large Binocular Telescope Observatory, 933 North Cherry Avenue, Tucson, AZ 85721, USA
                       \and
             Steward Observatory, Department of Astronomy, University of Arizona, 993 N. Cherry Avenue, Tucson, AZ 85721, USA
                       \and
             Finnish Centre for Astronomy with ESO (FINCA), FI-20014 University of Turku, Finland
                       \and
             Astronomy Department and Van Vleck Observatory, Wesleyan University, Middletown, CT 06459, USA
                       \and
             Department of Astronomy, University of California, Berkeley, CA, USA
                       \and
             ESA--ESAC, Operations Department, Camino Bajo del Castillo s/n, 28692 Villanueva de la Ca\~nada, Madrid, Spain
                       \and
             Aurora Technology B.V. for ESA, ESA--ESAC, Camino Bajo del Castillo s/n, 28692 Villanueva de la Ca\~nada, Madrid, Spain
                       \and
             Observatorio Astron\'omico Nacional (OAN-IGN) -- Observatorio de Madrid, Alfonso XII, 3, 28014 Madrid, Spain
                       \and
             INAF - Osservatorio Astronomico di Palermo Piazza del Parlamento 1, 90134 Palermo, Italy
                       \and
             Max-Planck-Institut f\"ur Astronomie, K\"onigstuhl 17, D-69117 Heidelberg, Germany
             }

   \date{Received June 2019; accepted \dots}

 
  \abstract
  {This paper is framed within a large project devoted to studying the presence of
  circumstellar material around main sequence stars, and looking for exocometary
  events. The work concentrates on HR 10 (A2 IV/V), known for its conspicuous
  variability  in the circumstellar narrow  absorption features of Ca {\sc ii} K
  and other lines, so far interpreted as $\beta$ Pic-like phenomena, within the
  falling evaporating body scenario.}
  {The main goal of this paper is to carry out a thorough study of HR 10 
  to find the origin of the observed variability, determine
  the nature of the star, its absolute parameters, and evolutionary status.}
  {Interferometric near-infrared (NIR) observations, multi-epoch high-resolution 
  optical spectra spanning a time baseline of more than 32 years, and optical and
  NIR photometry, together with theoretical modelling, were used to 
  tackle the above objectives.}
  {Our results reveal that HR 10 is a binary. The narrow 
  circumstellar absorption features superimposed on the photospheric Ca {\sc ii} K
  lines -- and lines of other species -- can be decomposed into two or more
  components, the two deep ones tracing the radial velocity of the individual
  stars, which implies  that their origin cannot be ascribed to transient
  exocometary events, their variability  being fully explained by the binarity 
  of the object. There does not appear to be transient events associated with
  potential exocomets. Each individual star holds its own circumstellar 
  shell and there are no traces of a circumbinary envelope. Finally, the
  combined use of the  interferometric and radial velocity data leads to a
  complete spectrometric and orbital solution for the binary, the main parameters
  being: an orbital period of 747.6 days, eccentricities of the orbits around the centre
  of mass 0.25 (HR 10-A), 0.21 (HR 10-B) and a mass ratio of $q\!=M_{\rm B}/M_{\rm
  A}\!=0.72-0.84$. The stars are slightly off the main sequence, the
  binary being $\sim\!530$ Myr old.}
  {}
  
   \keywords{Stars: binaries -- Stars: circumstellar matter -- Stars: fundamental
   parameters -- Line: profiles -- Techniques: interferometric -- Techniques: spectroscopic}
   
   \authorrunning{B. Montesinos et al.}
   \titlerunning{HR 10: A MS binary with two circumstellar shells}
   
   \maketitle
   
%

\section{Introduction}
\label{Sect:INTRODUCTION}

Once a star has undergone the pre-main sequence phase and the protoplanetary
-- primordial -- disc evolves, experiencing gas dispersal and grain 
growth \citep[see e.g.][]{Armitage15}, the surrounding circumstellar (CS
hereafter) material, namely comets, minor bodies, dusty debris discs, and 
gas in shells or envelopes, is as important as planets for providing clues to 
the early history of the solar system, and by analogy, of exoplanetary
systems. 

While exoplanets \citep[see e.g.][and references therein]{Beuther14} 
and debris discs \citep[see e.g.][]{Matthews14} are now routinely studied, little is
known about comets and other minor bodies around other stars. Their own 
nature, that is their small size and lack of an atmosphere, makes them elusive 
to any direct detection, but indirect methods, such as the detection of 
certain features originated by dust and gas, provide evidence of and 
information about the presence and properties of this kind of solid-phase 
CS material.

Dust features offer hints about the properties of $\mu$m-sized grains
in debris discs that result from collisions of planetesimals
\citep[e.g.][]{Olofsson12}. Circumstellar CO emission around some
AF-type main sequence (MS) stars \citep[e.g.][]{Moor15,Marino16} has been
interpreted as being the result of outgassing produced by comet collisions
\citep{Zuckerman12}. A cometary origin of warm \citep{Mennesson14,Ertel18} and
hot \citep{Absil06,Absil13,Ertel14}, exozodiacal dust has also been suggested
\citep{Faramaz17}.

On the other hand, spectroscopic monitoring of transient
absorption events in CS metallic lines -- the Ca {\sc ii} K line being
particularly sensitive and prominent \citep{Welsh15} -- can 
provide more direct information on the dynamics and composition of exocomets. 
The best-known examples are the variable absorption 
features observed in the spectra of \object{$\beta$ Pic} \citep{Hobbs85}, 
that are interpreted as the result of gas being released by the sublimation
of exocomets grazing or falling onto the star \citep[][and references
  therein]{Ferlet87,Kiefer14}, driven into its vicinity
by the perturbing action of a larger body, for example a planet. This
scenario is the so-called falling evaporating body model
\citep[FEB;][]{Beust90,Beust91}. Variable absorption features of this kind
have also been observed toward several A-type stars
\citep[e.g.][]{Redfield07,Roberge08,Welsh15}; another example is
the discovery and analysis by \citet{Grady18} of infalling gas in
\object{HD 172555}, in ultraviolet resonant transitions of Si {\sc iii} and {\sc
  iv}, C {\sc ii} and {\sc iv}, and O {\sc i}.

Since September 2015 we have been carrying out a high-resolution spectroscopic
project aiming at detecting and monitoring narrow, variable metallic
absorption lines that are attributed to transient exocometary events
around MS stars, mainly of A-type. Around 1600 spectra of more than 100
stars have been obtained \citep{Rebollido19}. 
The first results of this project are the study of the variations of the CS 
Ca {\sc ii} HK lines of the A5 V star \object{$\varphi$ Leo} \citep{Eiroa16}, 
and the co-existence of hot and cold gas in debris discs \citep{Rebollido18}.

In the context of this project, and in order to contribute to the 
understanding of the properties of CS material around MS or early
post-MS stars, we have focused this work on \object{HR 10} (A2 IV/V),   
a shell star showing very interesting spectroscopic behaviour 
linked to its CS material (see Sect. \ref{Sect:PREVIOUS_WORK}). The 
star has been treated so far as a single object; this fact has 
biased the interpretation of its observed spectroscopic variability 
towards scenarios that must be totally revised after the discovery,  
reported in this paper, that HR 10 is actually a binary.

In Sect. \ref{Sect:PREVIOUS_WORK} we summarise what is known about this
star and relevant to this work; in Sect. \ref{Sect:PIONIER} 
 we describe and analyse the PIONIER/VLTI observations that 
lead to the discovery that HR 10 is a binary; in Sect. \ref{Sect:SPECTRA} we
describe the high-resolution spectroscopic observations, both from
our own campaigns and from archives or the literature, that have been used to 
characterise the properties and measure the radial velocities (RVs) 
of the CS narrow absorption features superimposed on the photospheric
lines. Section \ref{Sect:CS_FEATURES} shows details and properties of those CS 
absorption features. In Sect. \ref{Sect:RESULTS} we give quantitative details
about the RVs, the orbital solution from the interferometric and spectroscopic 
RV data, and derive the stellar parameters of the individual components of 
HR 10. In Sect. \ref{Sect:DISCUSSION} we present a discussion of the 
results. Finally, in Sect. \ref{Sect:CONCLUSIONS} we summarise the 
main conclusions of this work.

\begin{table*}[!htb]
\begin{small}
\caption{PIONIER observations of HR\,10.}
\label{Table:PIONIER}
\setlength{\tabcolsep}{12pt}
\begin{tabular}{lcccc}
  \hline\hline
  \noalign{\smallskip}
Date                       & 2014-09-02     & 2017-09-23    & 2018-08-07    & 2018-09-05     \\
Starting time [UT]          & 06:13:06       & 02:55:38      & 09:15:56      & 07:50:37       \\ 
BJD [TDB] mid observation  & 2456902.7803   & 2458019.6773  & 2458337.8991  & 2458366.8422   \\
Programme                & 093.C-0712(B)   &  099.C-2015(A) & 0101.C-0182(B)& 0101.C-0182(B) \\
  \noalign{\smallskip}
  \hline
  \noalign{\smallskip} 
VLTI array                 & D0-G1-H0-I1    & A0-B2-C1-D0   & A0-G1-J2-J3   & A0-G1-J3-K0    \\
\# SCI observations$^a$    & 3              & 4             & 2             & 2              \\
Significance in $cp$       & 1435           & 50            & 283           & 247            \\
Significance in $V^2$      & 8271           & 1378          & 1623          & 4948           \\
Significance combined      & 1038           & 963           & 1046          & 654            \\
Separation [mas]           & 15.98$\pm$0.35 & 2.94$\pm$0.56 & 6.34$\pm$0.16 & 11.43$\pm$0.27 \\
Position angle [deg]$^b$   & 39.3$\pm$1.2   & 77.7$\pm$2.5  & 32.1$\pm$0.3  & 37.0$\pm$2.7   \\
Contrast (\%)              & 32.0$\pm$1.5   & 35$\pm$13     & 31.9$\pm$3.9  & 31.9$\pm$2.8   \\
\noalign{\smallskip}
\hline
\noalign{\smallskip}
\multicolumn{5}{l}{Notes: $(a)$ Number of observations (observing blocks) executed on the science target and bracketed by observations of}\\
\multicolumn{5}{l}{calibrators. $(b)$ The position angle is measured east of north, i.e. counterclockwise from the north.}
\end{tabular}
\end{small}
\end{table*}

\section{Previous work on HR 10}
\label{Sect:PREVIOUS_WORK}

HR 10 (HD 256, HIP 602) is a bright ``shell star'' ($V\!=\!6.23$) 
\citep{Cheng91,Jaschek91}, also labelled as a ``disc star'' by \citet{Abt08} 
under the assumption that due to the large rotation velocity, the CS Ti {\sc ii}
lines must originate in a disc. The spectral type classification of the object 
in the literature varies from A0 to A6, and luminosity classes IV or V. 
The current most accepted classification is A2 IV/V \citep[see e.g.][VizieR catalogue
  \texttt{III/231}]{Wright03}.  The spectrum of HR 10 resembles that
of a rapidly rotating early A star \citep[$v \sin i$=294 km/s,][]{Mora01}.

Observations taken since the late 1980s show high spectroscopic
variability in the narrow absorption components superimposed on the
photospheric Ca {\sc ii} K line. \citet{Lagrange-Henri90} and
\citet{Welsh98} interpreted those variations within the FEB model,
which is similar to that devised for $\beta$ Pic.

\citet{Redfield07} carried out a detailed study of the presence of 
narrow components superimposed on the photospheric Ca {\sc ii} and 
Na {\sc i} lines, and their potential short- and long-term variability. 
The origin of the narrow Ca {\sc ii} K absorption was attributed to CS material, 
with no interstellar contribution, the variability being caused by the 
closeness of the gas to the star. No significant absorption was detected 
in Na {\sc i} after water vapour lines were removed.

\citet{Abt08} pointed out the singularity of HR 10 as the only star in
a sample of 20 objects analysed showing double Ti {\sc ii} narrow absorptions 
in the doublet around 376.0 nm. The redshifted components were interpreted
as being caused by infalling material onto the star.

Concerning the spectral energy distribution (SED), \citet{Cheng91} 
reported a weak IR excess at 12 and 25 $\mu$m from Infrared Astronomical 
Satellite ({\em IRAS}) data, whereas \citet{Redfield07} did not detect any 
significant IR excess in {\em Spitzer} Infrared Array Camera (IRAC), Multiband
Imaging Photometer (MIPS), or Infrared Spectrograph (IRS) measurements. An 
upper limit of $L_{\rm IR}/L_*\!<\!6.1\times\!10^{-6}$ (consistent with the {\em
  Spitzer} upper limits at the longest IR wavelengths) was given by
these latter authors, a value to be compared with $L_{\rm
  IR}/L_*\!\simeq\!3.0\times\!10^{-3}$ \citep{Backman93}, or
$\simeq\!2.4\times\!10^{-3}$ \citep{Heinrichsen99} for $\beta$ Pic.
There are no traces of accretion or emission lines in the
spectra, and therefore the star is  most likely not in the pre-MS phase.

So far, HR 10 has been treated in the literature as a single
object. In retrospect, this fact has been a tight constraint not only
in interpreting the observed variability, but also in determining its
absolute parameters and its position in the HR diagram. The
single-star assumption yielded some controversial results; the $V$
magnitude, and the {\em Gaia} DR2 distance, $\varpi\!=\!6.8882\pm0.1184$ mas,
$d\!=\!145.18^{+2.54}_{-2.45}$ pc \citep{Gaia18}, respectively, imply an absolute magnitude of
$M_V\!\simeq\!+0.42$, which would put the star closer to luminosity class 
$\sim$III/IV; \citet{Redfield07} found a value $L_*/L_\odot\!=\!63.8$ which
corresponds to a typical $\sim$A2 III \citep{LB82}, both results being
in contrast with the spectral classification
IV/V. \citet{Montesinos09} estimated the stellar gravity from the
width of the wings of the Balmer lines and derived the mass and age from 
a $\log g_*$-- $\log T_{\rm eff}$ HR diagram and evolutionary tracks, and then
the luminosity from the corresponding point in the $\log
L_*/L_\odot$--$\log T_{\rm eff}$ HR diagram; the fact that the Balmer
lines -- and the whole spectrum -- are actually the composite of
profiles from two stars invalidates the initial measurement of $\log
g_*$ and then the set of parameters derived. The discovery of the
binarity of HR 10 makes a complete reassessment of the stellar 
parameters compulsory.

\begin{figure*}[!tbh]
\centering
\includegraphics[width=1.0\textwidth]{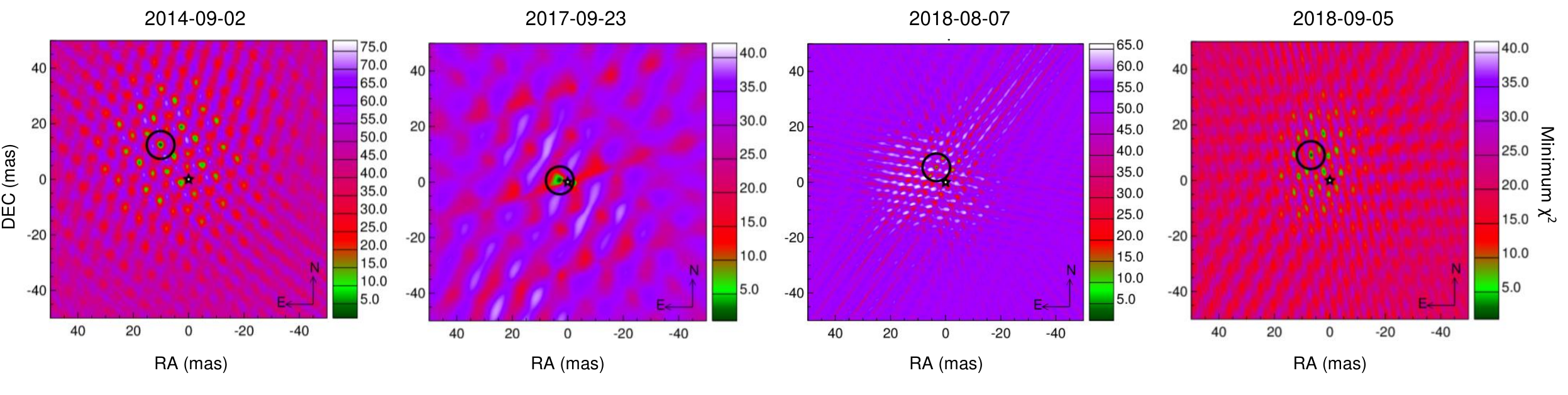}
\caption{Normalised $\chi$-square maps of the combined $cp\!+\!V^2$ 
for the four PIONIER/VLTI observations. The maps are centred at 
the brightest component. The black circles indicate the positions 
of the minima in the maps and the star marks the position of 
the bright component.}
\label{Fig:PIONIER_4panels}
\end{figure*}

\section{PIONIER/VLTI observations}
\label{Sect:PIONIER}

Interferometric observations of HR\,10 were first performed in September 2014
as part of a search for hot exozodiacal dust \citep[see details of the programme in][]{Ertel14, Ertel16} 
and three more times in September 2017, and August-September 2018,
for a total of four epochs. The Precision Integrated Optics Near-Infrared ExpeRiment
\citep[PIONIER,][]{LeBouquin11} on the Very Large Telescope
Interferometer (VLTI) was used. PIONIER is a four-telescope beam combiner
allowing the user to simultaneously obtain squared visibility ($V^2$)
measurements  on six baselines and closure phase ($cp$) measurements on
four telescope triplets.  A log of the observations is 
provided in Table~\ref{Table:PIONIER}.

PIONIER operates in $H$~band; the light was dispersed over three
(in 2014) and six spectral channels (later) across this band.  Chains
of alternating observations of a calibration star (CALs) and the
science target (SCI) were executed in order to characterise the
interferometric transfer function (TF) and calibrate the $V^2$ and
$cp$ measurements of the science target.  Each chain started and ended
with a CAL, and various CALs were used within a chain in order to
minimise the impact of imperfect knowledge of the calibration stars
(e.g. uncertain stellar diameters or unknown companions).

Data reduction and calibration were performed with the PIONIER
pipeline \texttt{pndrs} \citep{LeBouquin11} version~3.79 using
standard parameters.  In particular, we calibrated a whole sequence of
CAL and SCI observations at once using the smooth TF interpolation
method provided by the pipeline.  

The binary signature is obvious in both the calibrated and uncalibrated $V^2$ 
and $cp$ data, making for a strong detection with each observation.  
In order to extract the astrometric information (separation and position angle) 
and contrast (secondary-to-primary flux ratio) of the binary, we follow the method
outlined by \citet{Absil11} and \citet{Marion14}. We compute the $\chi^2$ 
goodness of fit to the $V^2$ and $cp$ data, both separately and jointly, for a series 
of binary star models with a range of positions and flux ratios. The resulting 
$\chi^2$ cubes (one for the $V^2$, one for the $cp$, and one for the combination 
of both) were used to identify the best-fit model and evaluate the significance of
the detection. The results are listed in Table \ref{Table:PIONIER}.
Figure \ref{Fig:PIONIER_4panels} shows the normalised $\chi$-square 
maps of the combined $cp\!+\!V^2$ for the four PIONIER/VLTI observations
of HR 10.

\begin{figure*}[!tbh]
\centering
\includegraphics[width=1.0\textwidth]{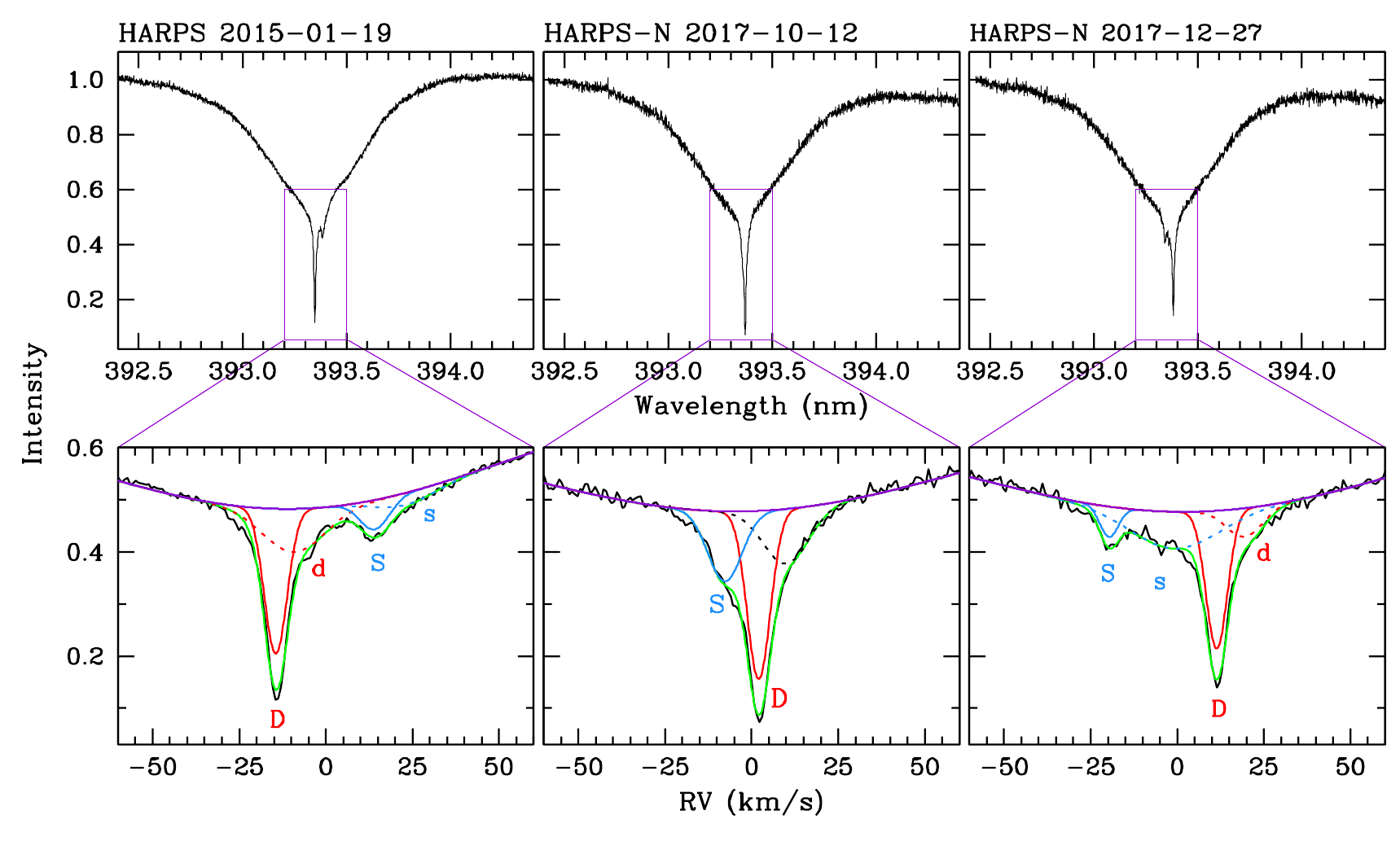}
\caption{Top: Profiles of the broad photospheric Ca {\sc ii} K
  line of HR 10 of spectra obtained on 19 Jan 2015, 12 Oct 2017 and
  27 Dec 2017. Superimposed on the lower part of the lines, narrow
  absorption features are remarkably conspicuous. Bottom: Narrow CS
  components corresponding to the spectra at the top, plotted in
  velocity space. The velocity ${\rm RV}\!=\!0$ km/s corresponds to the 
  rest wavelength of the Ca {\sc ii} K line. The letters D and S label the deep and
  shallow components of each CS absorption, respectively; these are present in all
  the spectra analysed in this work. Solid red and blue lines
  correspond to the Gaussian decomposition of those two main
  components, the bottom of the photospheric profile is plotted in
  purple and the whole fit in green. Dotted lines are the additional
  Gaussians d and s required to complete the decomposition of the absorption.
  See text for details.}
\label{Fig:Three_epochs}
\end{figure*}


\begin{table*}[!htbp]
\footnotesize
\setlength{\tabcolsep}{3.0pt}
\caption{Spectroscopic observations of HR 10.}
\label{Table:Spectra}

\begin{tabular}{lccclcccl}
\multicolumn{9}{l}{Dedicated spectroscopic campaigns.} \\ 
\hline\hline\noalign{\smallskip}
\multirow{2}{*}{Dates} && \multirow{2}{*}{N} &&\multirow{2}{*}{Resolution} && Spectral range  && Telescope / Observatory  \\
                       &&                    &&                            &&    (nm)         && Spectrograph             \\
\cline{1-1}\cline{3-3}\cline{5-5}\cline{7-7}\cline{9-9} \noalign{\smallskip}
2015-09\hspace{0.20cm} 04-07 && 17  &&            &&            && \multirow{2}{*}{1.2-m Mercator Telescope / La Palma}                                                 \\
2015-12\hspace{0.20cm} 20-23 && 8   && $85\,000$  && 377 -- 900 && \multirow{2}{*}{High Efficiency \& Resolution Mercator Echelle Spectrograph (HERMES)}                \\ 
2016-07\hspace{0.20cm} 12    && 1   &&            &&            &&                                                                                                      \\
\noalign{\vspace{0.1cm}}\hline\noalign{\vspace{0.1cm}}
\multirow{2}{*}{2015-10\hspace{0.20cm} 22-23} && \multirow{2}{*}{3} && \multirow{2}{*}{$48\,000$} && \multirow{2}{*}{350 -- 920} && 2.2-m ESO-MPIA Telescope / La Silla    \\
                                              &&                    &&                            &&      && Fibre-fed Extended Range Optical Echelle Spectrograph (FEROS)  \\
\noalign{\vspace{0.1cm}}\hline\noalign{\vspace{0.1cm}}
2017-10-05\hspace{0.15cm} to                  && \multirow{2}{*}{10} && \multirow{4}{*}{$67\,000$} && \multirow{4}{*}{370 -- 910 } &&                                                 \\
2018-01-03                                    &&                     &&                            &&                              && 2.5-m Nordic Optical Telescope (NOT) / La Palma \\
\noalign{\vspace{0.05cm}}
2018-11-23\hspace{0.15cm} to                  &&  \multirow{2}{*}{12}&&                            &&                              && FIbre-fed Echelle Spectrograph (FIES)           \\
2019-01-08                                    &&                     &&                            &&                              &&                                                 \\
\noalign{\vspace{0.1cm}}\hline\noalign{\vspace{0.1cm}}
2017-10-06\hspace{0.15cm} to                  && \multirow{2}{*}{21}  && \multirow{2}{*}{$115\,000$} && \multirow{2}{*}{383 -- 693} && 3.6-m Telescopio Nazionale Galileo (TNG) / La Palma \\ 
2018-01-02&&                      &&                          &&  &&        High Accuracy Radial velocity Planet Searcher North (HARPS-N)\\
\noalign{\vspace{0.05cm}}
\hline\noalign{\smallskip}
\multicolumn{9}{l}{Note: Dates of the specific nights of the campaigns are given in Table \ref{Table:RVs}} \\
\noalign{\vspace{0.4cm}}
\end{tabular}
\begin{tabular}{lccclccclcc}
\multicolumn{11}{l}{Observations extracted from publications or archives.}\\
\hline\hline\noalign{\smallskip}
\multirow{2}{*}{Dates} && \multirow{2}{*}{N} &&\multirow{2}{*}{Resolution} && Spectral range  && Telescope / Observatory && Ref.           \\
                       &&                    &&                            &&    (line or nm) && Spectrograph &&                           \\
\cline{1-1}\cline{3-3}\cline{5-5}\cline{7-7}\cline{9-9}\cline{11-11} \noalign{\smallskip}
1986-08-15 && 1  && $60\,000$                 && \multirow{2}{*}{Ca {\sc ii} K} && 1.4-m Coud\'e Auxiliary Telescope (CAT) / La Silla && \multirow{2}{*}{1}     \\
1998-10-15 && 1  && $100\,000$                &&                                && ESO Coud\'e Echelle Spectrometer (CES)             &&                        \\
\noalign{\vspace{0.1cm}}\hline\noalign{\vspace{0.1cm}}
1996-11-30 && 1  && \multirow{3}{*}{$1\,000\,000$}&& \multirow{3}{*}{Ca {\sc ii} K} && \multirow{2}{*}{3.6-m Telescope / Anglo Australian Observatory}    && 2  \\
1997-06-20 && 1  &&                               &&                                && \multirow{2}{*}{Ultra High Resolution Facility (UHRF)}             && 2  \\
2005-06-15 && 1  &&                               &&                                &&                                                                    && 3  \\
\noalign{\vspace{0.1cm}}\hline\noalign{\vspace{0.1cm}}
1998-10-25 && 1  && \multirow{2}{*}{$49\,000$}&& \multirow{2}{*}{380 -- 590}    && 4.2-m William Herschel Telescope / La Palma &&  \multirow{2}{*}{4} \\
1999-01-28 && 1  &&                           &&                                && Utrech Echelle Spectrograph (UES)           &&                     \\
\noalign{\vspace{0.1cm}}\hline\noalign{\vspace{0.1cm}}
2004-08-28 && 1  &&                           &&                   && \multirow{2}{*}{2.7-m Harlan J. Smith Telescope / McDonald Observatory}   &&    \\
2004-08-29 && 1  && $240\,000$                &&  Ca {\sc ii} K    && \multirow{2}{*}{Cross-Dispersed Echelle Spectrometer (CS21)}  &&  3 \\
2005-09-15 && 1  &&                           &&                   &&                                                               &&    \\
\noalign{\vspace{0.1cm}}\hline\noalign{\vspace{0.1cm}}
2007-06-28 && 14 && \multirow{5}{*}{$60\,000$} &&  \multirow{4}{*}{360 -- 450}&& \multirow{4}{*}{8-m ESO--Very Large Telescope / Paranal}               && \multirow{4}{*}{ESO}     \\
2007-07-11 &&  6 &&                            &&  \multirow{4}{*}{460 -- 670}&& \multirow{4}{*}{Ultraviolet and Visual Echelle Spectrograph (UVES)}    && \multirow{4}{*}{Archive} \\
2007-07-15 && 14 &&                            &&                                &&                                                                       &&          \\
2007-07-26 && 34 &&                            &&                                &&                                                                       &&          \\
2007-07-27 &&  6 &&                            &&                                &&                                                                       &&          \\
\noalign{\vspace{0.1cm}}\hline\noalign{\vspace{0.1cm}}
2010-09-01 && 3  &&                       && \multirow{2}{*}{336 -- 810}    && 10-m Keck I Telescope/ Hawaii                &&  Keck                   \\       
2011-12-08 && 3  &&                       &&                                && High Resolution Echelle Spectrometer (HIRES) &&  Archive                \\
\noalign{\vspace{0.1cm}}\hline\noalign{\vspace{0.1cm}}
2015-01-18 && 1 && \multirow{4}{*}{$120\,000$} && \multirow{4}{*}{378 -- 691}&& \multirow{3}{*}{3.6-m ESO Telescope / La Silla}  &&                                \\
2015-01-19 && 2 &&                        &&                                 && \multirow{3}{*}{High Accuracy Radial velocity Planet Searcher (HARPS)} &&  ESO     \\
2015-01-20 && 1 &&                        &&                                 &&                                                                        && Archive  \\
2015-01-21 && 2 &&                        &&                                 &&                                         &&                                         \\
\noalign{\vspace{0.1cm}}
\hline\noalign{\smallskip}
\multicolumn{11}{l}{Notes: ``N'' is the number of spectra. References: (1) \cite{Lagrange-Henri90}, (2) \cite{Welsh98},}\\
\multicolumn{11}{l}{\hspace{0.9cm} (3) \cite{Redfield07}, (4) EXPORT La Palma International Time campaigns, \cite{Mora01}.}\\
\end{tabular}
\end{table*} 


\section{Spectroscopic observations}
\label{Sect:SPECTRA}

In this section we describe the sets of high-resolution spectroscopic
observations of HR 10 that have been used in this work to
analyse the CS narrow absorption features superimposed on the photospheric
lines, decompose their complex profiles, and measure the corresponding
RVs of the components. We built a historical record covering the
period 1986--2019 in order to complete what is to date the most
comprehensive study of the CS variability of this object.

\subsection{Dedicated campaigns}

\noindent Table \ref{Table:Spectra} (upper block) summarises the
information of the campaigns coordinated by our team since September 2015 to
observe \mbox{HR\,10}, among other stars, looking for transient exocometary
events around MS stars, and campaigns targeted exclusively at HR
10. Signal-to-noise ratios (S/Ns) of the spectra at the bottom
of the Ca {\sc ii} K line were always above 100. The spectra used in
this work come from the corresponding reduction pipelines.

\subsection{Archival and published data}

In Table \ref{Table:Spectra} (lower block) we give the relevant
information for the observations obtained prior to September 2015, when we
started our own monitoring campaigns. The data come from
the literature, were provided by one of the authors (Prof. Seth Redfield), or 
were taken from observatory archives. 
Direct measurements on the reduced spectra, corrected for barycentric
velocity, were done for the observations obtained with UES/WHT,
CS21/2.7-m HJS, UVES/VLT, HIRES/Keck, and HARPS/3.6-m La Silla, and on
the spectrum taken on 15 Jun 2005 with UHRF/3.6-m AAO. The spectra from
\cite{Lagrange-Henri90} were not available in readable format,
and were therefore scanned from the paper itself and
digitized. The data on radial velocities from the observations obtained 
on 30 Nov 1996 and 20 Jun 1997 were taken directly from the work by
\cite{Welsh98}.

\section{The circumstellar absorption features}
\label{Sect:CS_FEATURES}

\subsection{The Ca {\sc ii} K line}

Figure \ref{Fig:Three_epochs} (top) shows the profiles of the broad
photospheric Ca {\sc ii} K line of spectra of HR 10 obtained on
19 Jan 2015, 12 Oct 2017, and 27 Dec 2017. Superimposed on the lower part
of the lines, narrow absorption features are remarkably conspicuous. The
insets delimited with purple boxes are blown up and plotted in velocity
space at the bottom row; ${\rm RV}\!=\!0$ km/s corresponds to the rest
wavelength of the Ca {\sc ii} K line, namely, 393.368 nm.

These three observations are representative of what we see in the {\em
whole set} of spectra analysed in this work. The decomposition of 
the profiles in Gaussian components was carried out using the 
\texttt{Emission Line Fitting (ELF)} package of \texttt{dipso}, a spectrum 
analysis program developed by the Starlink Project \citep{Howarth04}.
The common feature of all the narrow absorptions is the presence 
of two main components: one deep and one shallow, hereafter plotted as red 
and blue solid lines, respectively, when the absorption is decomposed into
Gaussian functions and labelled `D' and `S'. The origin of these
narrow absorption features is clearly CS, their variability being 
attributed in previous works to $\beta$ Pic-like phenomena (see Sect. 
\ref{Sect:PREVIOUS_WORK}), however the interpretation of their dynamical 
behaviour, as we see below, when analysing the RV series over a 
long time span turns out to be related to the binarity of HR 10 rather than to transient phenomena.

The profiles of the observations taken on 19 Jan 2015 and 27 Oct 2017
(left and right panels of Fig. \ref{Fig:Three_epochs}) show two
cases where the D and S components appear well separated, whereas the
observation taken on 12 Oct 2017, plotted in the middle, shows a case where the
separation in radial velocity between D and S is small. 
The pattern observed during the $\sim\!32$ years of observations
collected is that component D  moves to the blue (red), as 
component S moves to the red (blue), crossing each other at certain times. 

When the profile was such that the two main components appeared well separated,
a four-Gaussian fit was needed to reproduce the whole narrow absorption.
The two additional components, plotted as dotted red and blue lines, are placed 
to the red side of the main D and S components, respectively; these latter have been 
labelled as `d' and `s', to stress the fact that each one seems to be associated 
with D and S. When the profile showed an appearance similar to that plotted 
in the middle graph of Fig. \ref{Fig:Three_epochs}, a three-Gaussian fit was 
accurate enough to reproduce the absorption. Whereas in the four-Gaussian fittings 
the d and s components, redshifted with respect to D and S, seem to have a physical
relationship with their corresponding main component, in the case of three-Gaussian 
fits, the parameters of the additional component probably embed the contribution 
of the two dotted components of the four-Gaussian decomposition. In Fig.
\ref{Fig:Three_epochs} the pseudocontinuum tracing the bottom of the broad
photospheric line is plotted in purple and the fit to the whole profile is plotted 
in green. For the three-Gaussian fit the additional component has been plotted as 
a dotted black line.

Figure \ref{Fig:201710-201801} shows the results of a well-sampled
monitoring carried out between 5 Oct 2017 and 3 Jan 2018 (21 spectra
obtained with FIES/NOT and HARPS-N/TNG), where the gradual dynamical
evolution of D and d moving to the red, and S and s moving to the blue, 
is very clear. The colour code is the same as in Fig. \ref{Fig:Three_epochs}.

We quantitatively analyse the behaviour of all these components 
in Sects. \ref{Sect:RVs} and \ref{Sect:DSds}. 

\subsection{Circumstellar absorption features in other photospheric lines}

In this work we concentrate on the analysis of the Ca {\sc ii} K line
for two main reasons: the first one is that this line is the most
important and clear tracer of CS activity, as has been proved in
$\beta$ Pic and other stars (see references in the Introduction); the
second is that for some of the early spectroscopic observations,
which have been crucial for the completion this study, only data for this
particular line were available. However, HR 10 presents narrow CS
absorption features superimposed on many other photospheric lines.

\begin{figure*}
\centering
\includegraphics[width=0.8\textwidth]{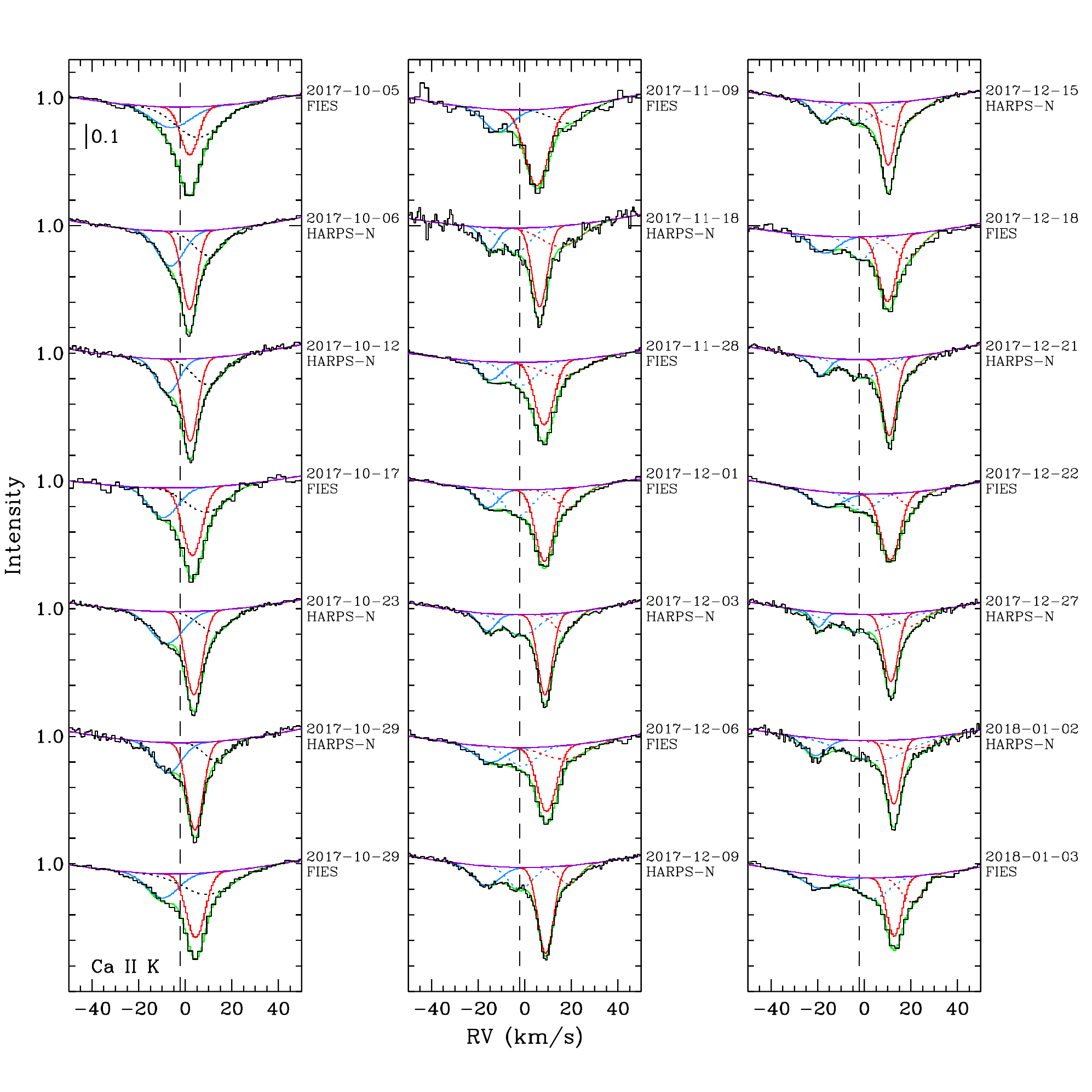}
\caption{The Ca {\sc ii} K CS profile for 21 observations obtained with 
FIES/NOT and HARPS-N/TNG between 5 Oct 2017 and 3 Jan 2018. The profiles have 
been decomposed with three or four Gaussians, plotting D (d) and S (s) as 
solid (dotted) red and blue lines. The pseudocontinuum tracing the bottom of 
the broad photospheric line is plotted in purple and the fit to the whole profile 
is plotted in green. For the three-Gaussian fit the additional component has been 
plotted as a dotted black line. The dashed line marks the RV of 
the system $-2.18\pm0.32$ km/s (see Sect. \ref{Sect:Spec_binary_sol} and Table \ref{Table:RV_res}).}
\label{Fig:201710-201801}
\end{figure*}

\begin{figure*}
\centering
\includegraphics[width=0.8\textwidth]{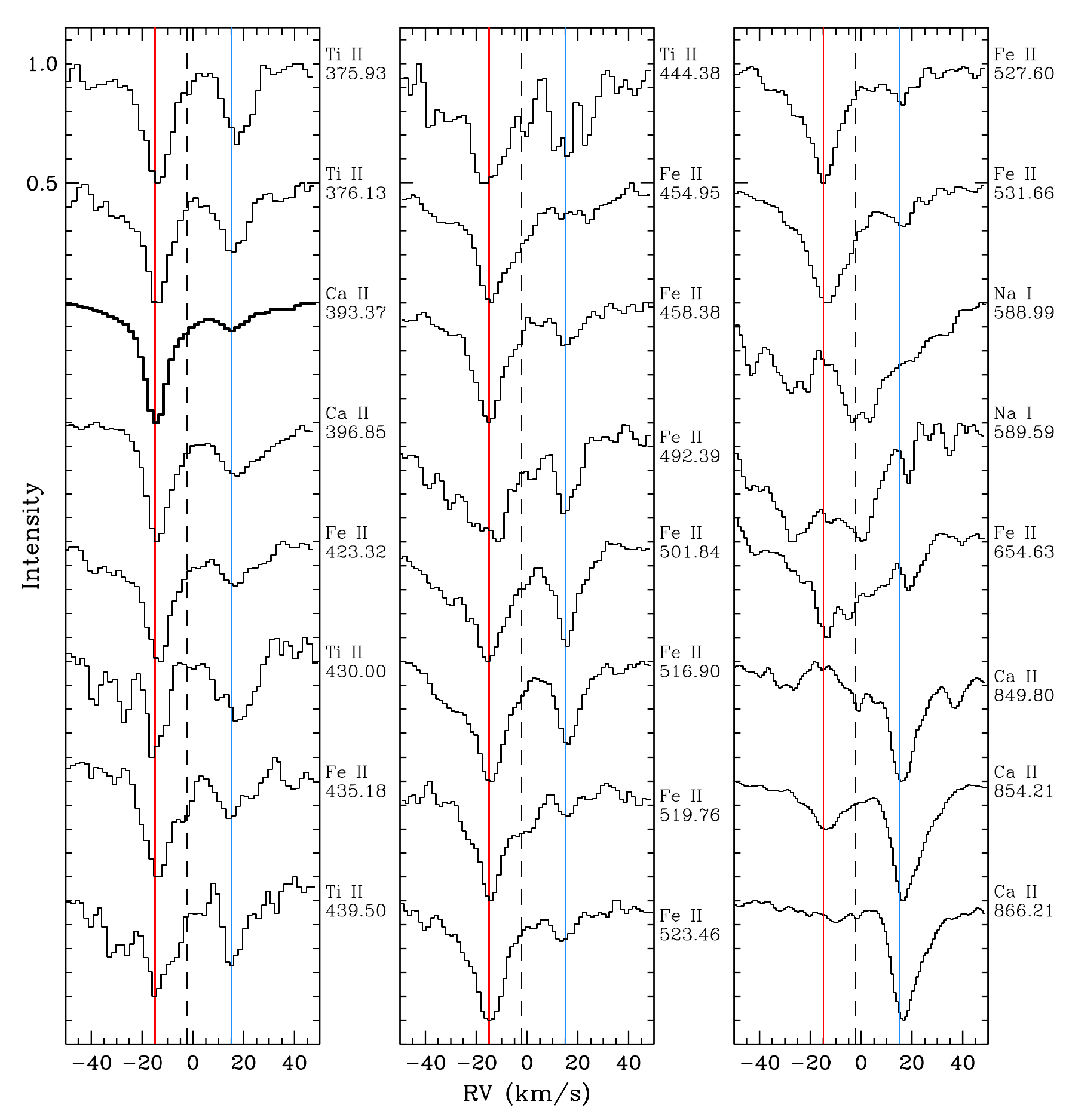}
\caption{Circumstellar components in 24 lines of the optical spectrum
  of HR 10. The photospheric profiles have been subtracted and, for the
  sake of clarity, the narrow absorption features have been scaled to the same
  size. The red and blue lines mark the radial velocities of the D
  (deep) and S (shallow) components, $-14.80\pm0.50$ $+15.16\pm1.54$
  km/s, respectively, of the Ca {\sc ii} K narrow absorption, which has
  been plotted with a thicker line. The dashed line marks the RV of the system, $-2.18\pm0.32$ km/s. The line identification
  with the wavelength in nm is included. See text for details.}
\label{Fig:CS_other_lines}
\end{figure*}

\begin{figure}
\centering
\includegraphics[width=0.35\textwidth]{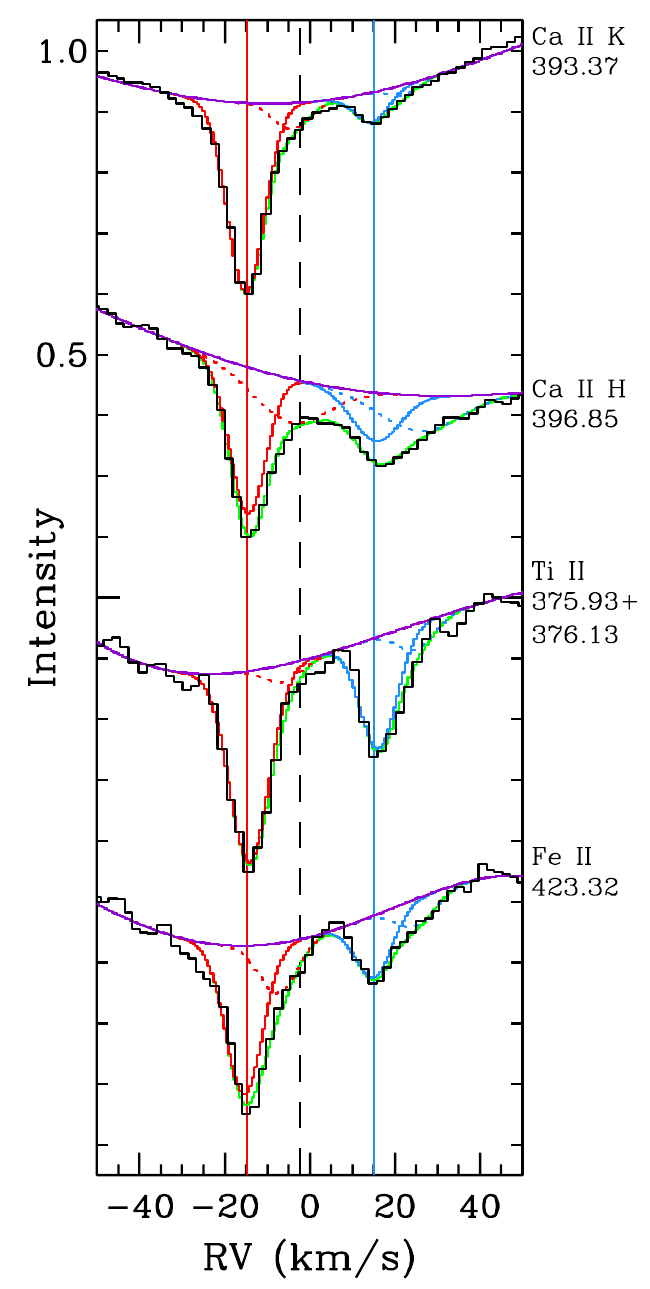}
\caption{Circumstellar components of the narrow absorption features superimposed 
to the photospheric Ca {\sc ii} K and H lines, the Ti {\sc ii} doublet around 
376.0 nm, and the Fe {\sc ii} line at 423.32 nm, for the same set of observations 
as in Fig. \ref{Fig:CS_other_lines}; the vertical solid red and blue lines and the
dashed black line mark the same radial velocities as in that figure. The 
Gaussian decomposition clearly shows the presence of the S, D 
(solid red and blue lines), s, and d (dotted red and blue lines) components.
The line identification with the wavelength in nm is included. 
See text for details.}
\label{Fig:CS_4lines4components}
\end{figure}

\begin{figure}
\centering
\includegraphics[width=0.35\textwidth]{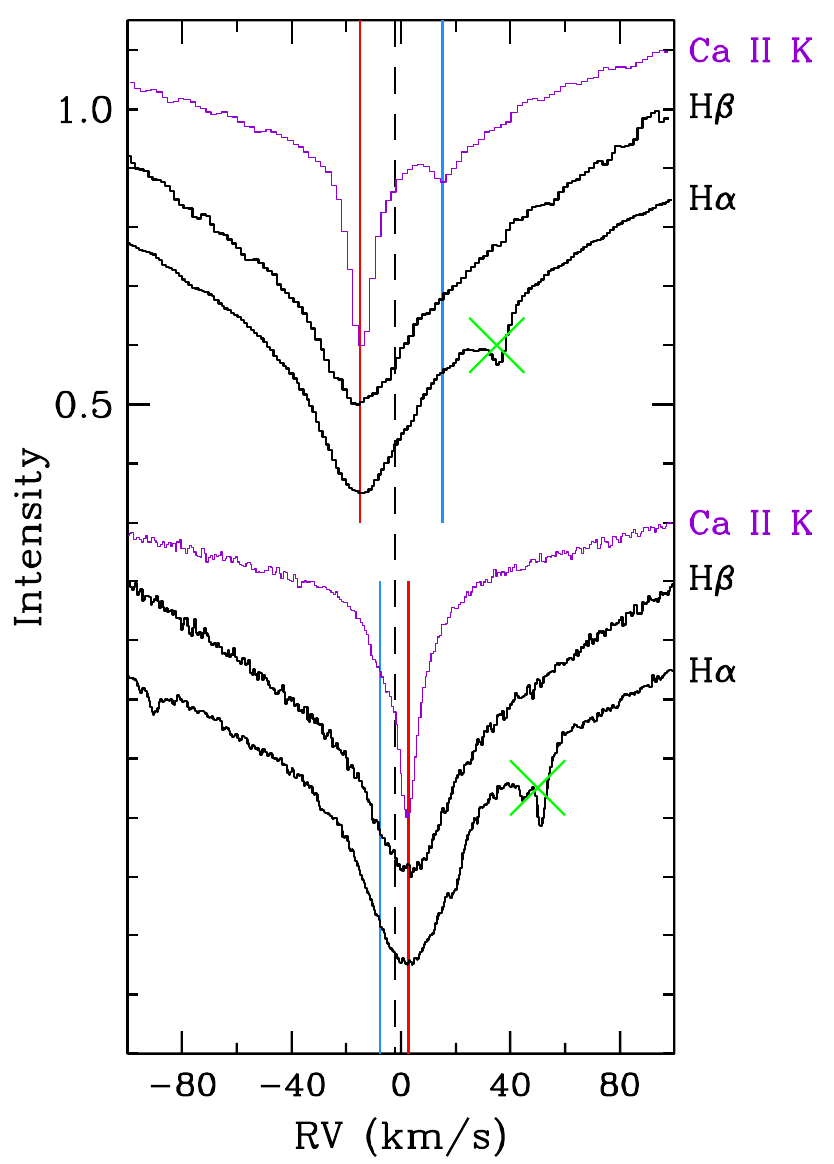}
\caption{Bottom of the Balmer lines H$\alpha$ and H$\beta$ for 
 observations corresponding to a configuration where the difference
 in RVs is large (top) and small (bottom). The red and blue lines mark 
 the radial velocities of the D and S components, measured on the Ca 
 {\sc ii} K narrow absorption features, plotted in purple as references. 
 Those RV are $-14.80\pm0.50$ (D), $+15.16\pm1.54$ km/s (S) (top), and
  $+2.75\pm0.73$ (D), $-7.72\pm1.38$ (S) km/s (bottom). The dashed line marks 
  the RV of the system, $-2.18\pm0.32$ km/s. The two green 
  crosses mark the position of telluric lines.}
\label{Fig:CS_Balmer}
\end{figure}

Figure \ref{Fig:CS_other_lines} shows the CS components superimposed on
24 photospheric lines of the spectrum of HR 10. Since some of these
components are fairly weak, in order to increase the S/N, five FIES/NOT spectra 
-- those taken between 26 Dec 2018 and 8 Jan 2019, see Table \ref{Table:RVs} -- 
were averaged. These were chosen because the separation in RVs of the 
Ca {\sc ii} K narrow absorption features is large, almost 30 km/s, and
therefore the CS components present in other lines would appear well separated. 
The photospheric profile of each line has been subtracted and the residual narrow
absorption features have been scaled between intensities 0.5 and 1.0. 
The red and blue solid lines are the average RVs of the D and S components 
of the Ca {\sc ii} K absorption, and the dashed line marks the RV of the 
system. The Ca {\sc ii} K CS absorption is plotted with a thicker line type.

In most cases, the CS narrow absorption features follow the same pattern of
intensity as in Ca {\sc ii} K, with the remarkable exception of the
IR Ca {\sc ii} triplet, where a strong component is seen in the
three lines at the velocity of the shallow component of the Ca {\sc
  ii} K absorption. Conversely, a weak absorption at the velocity of the
Ca {\sc ii} K strong component is clearly visible in the central and
most intense line of the triplet (854.21 nm), but does not appear in
the other two. On the other hand, the appearance of the absorption features 
at the Na {\sc i} D doublet is complex due to the many telluric 
lines in that spectral range; these were removed using the package
\texttt{MOLECFIT}\footnote{\url{http://www.eso.org/sci/software/pipelines/skytools/molecfit}}
\citep{Smette15,Kausch15}. 

Figure \ref{Fig:CS_4lines4components} shows the narrow CS components superimposed 
on the photospheric Ca {\sc ii} K and H lines, the Ti {\sc ii} 375.93 + 376.13 nm 
and the Fe {\sc ii} 423.32 nm lines; in the case of the Ti {\sc ii} doublet, the
profiles of the two CS absorptions of the two lines, in velocity space, were
merged and converted into a single profile. Colour codes and the meaning of the 
vertical lines are the same as in Figs. \ref{Fig:Three_epochs},
\ref{Fig:201710-201801}, and \ref{Fig:CS_other_lines}. It is remarkable that
components s and d, which appear in the Gaussian decomposition of the 
Ca {\sc ii} K narrow absorptions, are also present in other species.

Figure \ref{Fig:CS_Balmer} shows the bottom of the photospheric Balmer lines
H$\alpha$ and H$\beta$ for two situations with the D and S components 
of Ca {\sc ii} K -- plotted in purple as a reference -- appear well separated
(top) and blended (bottom). The profiles at the top correspond to the same 
spectra used in Fig. \ref{Fig:CS_other_lines}, whereas the profiles at the 
bottom are extracted from an average of the HARPS-N spectra obtained on 6 Oct 2017
and 24 Oct 2017. The vertical red and blue lines mark the RVs of components D and S 
 and the dashed black line marks the RV of the system; three 
telluric features on the H$\alpha$ profiles are marked by green crosses. 
It is very clear that the minima of the Balmer profiles are located 
at the RV of component D. The profiles at the top are clearly
asymmetric. In the following section it becomes clear why this happens, the binarity 
being the origin of the variable shape and position of the bottom of these lines.

We defer to a forthcoming paper a more detailed analysis of the CS absorption 
features in lines other than Ca {\sc ii} K, and also of the Balmer lines.

\section{Results}
\label{Sect:RESULTS}


\begin{table*}[!htbp]
\scriptsize
\setlength{\tabcolsep}{4.0pt}
\caption{Radial velocities and velocity dispersions ($b$) of the deep `D' and
shallow `S' components of the CS Ca {\sc ii} K absorptions. For those cases when a
four-Gaussian fit was feasible, the parameters of the weak `d' and `s' components 
that appear redshifted with respect to D and S, respectively, are also given.}
\label{Table:RVs}
\begin{tabular}{lc cc cc rrc rrc rrc rrc l}
\hline\hline\noalign{\smallskip}
                        &&                    &&           && \multicolumn{11}{c}{Component} && \\
\cline{7-17}\noalign{\smallskip}
\multirow{2}{*}{Date}   && \multirow{2}{*}{N} && BJD (TDB) && \multicolumn{2}{c}{D}   && \multicolumn{2}{c}{S}
                                                           && \multicolumn{2}{c}{d}  && \multicolumn{2}{c}{s}  &&   \multirow{2}{*}{Instrument}  \\
\cline{7-8}\cline{10-11}\cline{13-14}\cline{16-17}\noalign{\smallskip}
                        &&                    && (days)    && \multicolumn{1}{c}{RV} & \multicolumn{1}{c}{$b$} 
                                                           && \multicolumn{1}{c}{RV} & \multicolumn{1}{c}{$b$} 
                                                           && \multicolumn{1}{c}{RV} & \multicolumn{1}{c}{$b$} 
                                                           && \multicolumn{1}{c}{RV} & \multicolumn{1}{c}{$b$} && \\
                        &&                    &&           && \multicolumn{1}{c}{(km/s)} & \multicolumn{1}{c}{(km/s)} 
                                                           && \multicolumn{1}{c}{(km/s)} & \multicolumn{1}{c}{(km/s)}
                                                           && \multicolumn{1}{c}{(km/s)} & \multicolumn{1}{c}{(km/s)} 
                                                           && \multicolumn{1}{c}{(km/s)} & \multicolumn{1}{c}{(km/s)} &&  \\
\cline{1-1}\cline{3-3}\cline{5-5}\cline{7-8}\cline{10-11}\cline{13-14}\cline{16-17} \cline{19-19}\noalign{\smallskip}                       
1986-08-15 &&  1                   &&   2446657.5000 &&  $ -8.66\pm2.00$ & 10.1 &&  $ +9.38\pm2.00$ &  9.0 && & && & && CES     \\
1988-10-15 &&  1                   &&   2447449.5000 &&  $ -4.49\pm2.00$ &  9.8 &&  $ +5.83\pm2.00$ & 10.3 && & && & && CES     \\
1996-11-30($\dagger$) &&  1        &&   2450417.5022 &&  $-10.40\pm2.00$ &  2.1 &&  $ +7.40\pm2.00$ &  3.8 && & && & && UHRF    \\
1997-06-20($\dagger$) &&  1        &&   2450619.5011 &&  $ +9.20\pm2.00$ &  3.0 &&  $-16.60\pm2.00$ &  2.6 && $+16.20\pm2.00$ &  4.5 && & && UHRF \\
1998-10-25 &&  1                   &&   2451112.4652 &&  $-16.03\pm2.00$ & 10.0 &&  $+10.00\pm2.00$ & 10.0 && $-6.51\pm3.00$ & 13.7 && $+20.01\pm3.00$ & 9.8 && UES     \\
1999-01-28 &&  1                   &&   2451207.3363 &&  $ -8.78\pm2.00$ & 10.0 &&                  &      && & && & && UES     \\
2004-08-28 &&  \multirow{2}{*}{2}  &&  \multirow{2}{*}{2453246.3741} && \multirow{2}{*}{$-17.33\pm2.00$} & \multirow{2}{*}{8.5:}
                                                                     && \multirow{2}{*}{$+13.80\pm2.00$} & \multirow{2}{*}{7.9:} && &  && &  &&  \multirow{2}{*}{CS21} \\
2004-08-29 &&                      &&                &&                  &      &&                  &      && & && & &&         \\                    
2004-10-19 &&  1                   &&   2453297.7801 &&  $-14.77\pm2.00$ &      &&                  &      && & && & && CS21    \\
2005-06-15 &&  1                   &&   2453537.3058 &&  $ -2.10\pm2.00$ & 12.3:&&                  &      && & && & && UHRF    \\
2005-09-15 &&  1                   &&   2453628.8717 &&  $ +9.32\pm2.00$ & 14.7 &&  $-17.38\pm2.00$ & 16.3:&& &                       && $ -4.43\pm2.00$ &  7.4   && CS21    \\
2007-06-28 &&  14                  &&   2454279.8221 &&  $ -0.59\pm1.50$ &  7.8 &&  $ -4.40\pm1.50$ & 7.0  && & && & && UVES    \\
2007-07-11 &&  6                   &&   2454292.8971 &&  $ +0.31\pm1.50$ &  6.2 &&  $ -6.04\pm1.50$ & 4.6  && & && & && UVES    \\
2007-07-15 &&  14                  &&   2454296.9144 &&  $ +0.91\pm1.50$ &  6.4 &&  $ -6.07\pm1.50$ & 6.0  && & && & && UVES    \\
2007-07-26 &&  34                  &&   2454307.7830 &&  $ +1.27\pm1.50$ &  7.5 &&  $ -8.96\pm1.50$ & 8.0  && & && & && UVES    \\
2007-07-27 &&  6                   &&   2454308.7539 &&  $ +2.18\pm1.50$ &  8.0 &&  $ -7.82\pm1.50$ & 8.0  && & && & && UVES    \\
2010-09-01 &&  1                   &&   2455441.1252 &&  $-14.33\pm1.50$ & 10.5 &&  $+14.05\pm1.50$ & 9.8  && $ -4.01\pm2.00$ &  8.4  && $+18.73\pm2.00$ & 10.4   && HIRES   \\
2011-12-08 &&  1                   &&   2455903.7969 &&  $+15.17\pm1.50$ &  8.0 &&  $-24.10\pm1.50$ & 10.4 && $+23.67\pm2.00$ & 10.2  && $ -4.97\pm2.00$ &  8.0   && HIRES   \\
2015-01-18 &&  1                   &&   2457041.4530 &&  $-14.41\pm1.50$ &  8.2 &&  $+14.05\pm1.50$ & 11.0 && $ -4.00\pm2.00$ &  9.4  && $+24.10\pm2.00$ &  9.8   && HARPS   \\
2015-01-19 &&  2                   &&   2457042.4529 &&  $-14.45\pm1.50$ &  8.0 &&  $+14.15\pm1.50$ & 11.4 && $ -6.24\pm2.00$ & 10.2  && $+23.30\pm2.00$ & 10.0   && HARPS   \\
2015-01-20 &&  1                   &&   2457043.4527 &&  $-14.41\pm1.50$ &  8.8 &&  $+13.65\pm1.50$ & 11.0 && $ -1.99\pm2.00$ & 11.0  && $+23.28\pm2.00$ &  9.7   && HARPS   \\
2015-01-21 &&  2                   &&   2457044.4527 &&  $-14.58\pm1.50$ &  8.4 &&  $+14.22\pm1.50$ & 11.0 && $ -3.99\pm2.00$ & 10.3  && $+23.03\pm2.00$ & 11.0   && HARPS   \\
2015-09-04 &&  \multirow{2}{*}{17} &&   \multirow{2}{*}{2457271.0769} && \multirow{2}{*}{$+0.08\pm1.50$} &  \multirow{2}{*}{9.4}
                                                                      && \multirow{2}{*}{$-7.36\pm1.50$} & \multirow{2}{*}{13.1}
                                                                      && \multirow{2}{*}{$+10.31\pm2.00$}& \multirow{2}{*}{15.8}
                                                                      && \multirow{2}{*}{$+5.95\pm2.00$} & \multirow{2}{*}{11.1} && \multirow{2}{*}{HERMES} \\
2015-09-07 &&                      &&                &&                  &  &&                   & &&                &  &&                & &&          \\
2015-10-22 &&  \multirow{2}{*}{3}  &&   \multirow{2}{*}{2457318.1488} && \multirow{2}{*}{$+5.41\pm1.50$} &  \multirow{2}{*}{9.2}
                                                                      && \multirow{2}{*}{$-14.88\pm1.50$}&  \multirow{2}{*}{9.4}
                                                                      && \multirow{2}{*}{$+14.78\pm2.00$}& \multirow{2}{*}{19.9:}
                                                                      && \multirow{2}{*}{$-3.22\pm2.00$} & \multirow{2}{*}{19.7:} && \multirow{2}{*}{FEROS} \\
2015-10-23 &&                      &&                &&                  &  &&                   & &&                &  &&                & &&          \\ 
2015-12-20 &&  3                   &&   2457377.3295 &&  $+12.22\pm1.50$  & 10.5 &&  $-19.22\pm1.50$  & 11.9 && $+24.37\pm2.00$ & 10.2 && $-3.02\pm2.00$ & 13.0 && HERMES  \\
2015-12-22 &&  2                   &&   2457379.3572 &&  $+13.03\pm1.50$  &  9.5 &&  $-22.19\pm1.50$  & 13.6 && $+25.85\pm2.00$ & 11.0 && $-4.11\pm2.00$ &  8.8 && HERMES  \\
2015-12-23 &&  3                   &&   2457380.3501 &&  $+13.12\pm1.50$  &  9.6 &&  $-21.28\pm1.50$  & 10.0 && $+27.98\pm2.00$ & 11.0 && $-5.80\pm2.00$ & 14.0 && HERMES  \\
2016-07-12 &&  1                   &&   2457581.7080 &&  $ -0.63\pm1.50$  &  7.1 &&  $ -5.95\pm1.50$  &  6.1 && &  && &   && HERMES  \\
2017-10-05 &&  1                   &&   2458031.5276 &&  $ +1.84\pm1.50$  &  9.0 &&  $ -6.05\pm1.50$  & 19.4 && &  && &   && FIES    \\
2017-10-06 &&  1                   &&   2458032.5772 &&  $ +1.80\pm1.50$  &  7.7 &&  $ -6.36\pm1.50$  & 13.3 && &  && &   && HARPS-N \\
2017-10-12 &&  2                   &&   2458038.5469 &&  $ +2.07\pm1.50$  &  7.9 &&  $ -7.69\pm1.50$  & 11.5 && &  && &   && HARPS-N \\
2017-10-17 &&  1                   &&   2458043.5495 &&  $ +3.00\pm1.50$  &  9.9 &&  $ -9.46\pm1.50$  & 14.5 && &  && &   && FIES    \\
2017-10-23 &&  \multirow{2}{*}{2}  &&   \multirow{2}{*}{2458050.4980} && \multirow{2}{*}{$+3.18\pm1.50$} &  \multirow{2}{*}{4.8}
                                                                      && \multirow{2}{*}{$-9.12\pm1.50$} & \multirow{2}{*}{13.2}
                                                                      && \multirow{2}{*}{$+15.99\pm2.00$}& \multirow{2}{*}{15.7}
                                                                      && \multirow{2}{*}{$+4.14\pm2.00$} & \multirow{2}{*}{12.1} &&  \multirow{2}{*}{HARPS-N}\\
2017-10-24 &&                      &&                &&                   &      &&                   &      &&                 &      &&                 &      &&         \\
2017-10-29 &&  1                   &&   2458056.3919 &&  $ +4.40\pm1.50$  &  9.5 &&  $ -9.74\pm1.50$  & 15.5 &&                 &      &&                 &      && FIES    \\
2017-10-29 &&  2                   &&   2458056.4748 &&  $ +4.26\pm1.50$  &  5.4 &&  $ -9.57\pm1.50$  & 17.0:&& $+16.02\pm2.00$ & 20.5:&& $ +4.47\pm2.00$ & 16.1:&& HARPS-N \\
2017-11-09 &&  1                   &&   2458067.4778 &&  $ +5.15\pm1.50$  & 11.9 &&  $-11.31\pm1.50$  & 13.9 &&                 &      &&                 &      && FIES    \\
2017-11-18 &&  2                   &&   2458076.3180 &&  $ +6.38\pm1.50$  &  9.6 &&  $-15.05\pm1.50$  & 13.2 && $+20.01\pm2.00$ & 14.1 && $ -5.61\pm2.00$ & 10.0 && HARPS-N \\
2017-11-28 &&  1                   &&   2458086.4462 &&  $ +8.32\pm1.50$  & 10.4 &&  $-14.88\pm1.50$  & 11.4 && $+20.62\pm2.00$ & 10.0 && $ -2.61\pm2.00$ & 12.1 && FIES    \\
2017-12-01 &&  1                   &&   2458089.4417 &&  $ +8.48\pm1.50$  &  9.2 &&  $-15.91\pm1.50$  & 10.5 && $+18.97\pm2.00$ & 13.2 && $ -2.84\pm2.00$ & 13.8 && FIES    \\
2017-12-03 &&  2                   &&   2458091.3887 &&  $ +8.71\pm1.50$  &  9.1 &&  $-15.84\pm1.50$  & 11.0 && $+20.45\pm2.00$ & 11.3 && $ -3.97\pm2.00$ &  9.6 && HARPS-N \\
2017-12-06 &&  1                   &&   2458094.4387 &&  $ +9.46\pm1.50$  & 10.9 &&  $-15.10\pm1.50$  & 12.2 && $+23.01\pm2.00$ & 10.0 && $ -3.00\pm2.00$ & 10.9 && FIES    \\
2017-12-09 &&  2                   &&   2458097.3047 &&  $ +8.99\pm1.50$  &  7.9 &&  $-16.55\pm1.50$  & 13.6 && $+19.03\pm2.00$ & 10.4 && $ -1.95\pm2.00$ & 11.8 && HARPS-N \\
2017-12-15 &&  2                   &&   2458103.3163 &&  $+10.29\pm1.50$  &  9.5 &&  $-17.72\pm1.50$  & 10.8 && $+22.45\pm2.00$ & 11.0 && $ -3.09\pm2.00$ & 11.5 && HARPS-N \\
2017-12-18 &&  1                   &&   2458106.3321 &&  $+10.02\pm1.50$  &  9.0 &&  $-17.03\pm1.50$  & 12.9 && $+18.61\pm2.00$ & 13.1 && $ -1.62\pm2.00$ & 11.4 && FIES    \\
2017-12-21 &&  2                   &&   2458109.3041 &&  $+10.67\pm1.50$  &  9.0 &&  $-19.01\pm1.50$  & 11.6 && $+21.65\pm2.00$ &  7.0 && $ -4.00\pm2.00$ &  8.5 && HARPS-N \\
2017-12-22 &&  1                   &&   2458110.3674 &&  $+11.18\pm1.50$  &  9.8 &&  $-16.14\pm1.50$  & 15.8 && $+20.87\pm2.00$ &  9.8 && $ -0.36\pm2.00$ & 12.3 && FIES    \\
2017-12-27 &&  2                   &&   2458115.3078 &&  $+11.39\pm1.50$  &  7.2 &&  $-19.69\pm1.50$  & 11.2 && $+20.05\pm2.00$ & 12.2 && $ -6.82\pm2.00$ &  7.0 && HARPS-N \\
2018-01-02 &&  2                   &&   2458121.3055 &&  $+12.71\pm1.50$  &  7.3 &&  $-21.15\pm1.50$  & 12.1 && $+21.05\pm2.00$ & 10.1 && $ -2.91\pm2.00$ &  9.0 && HARPS-N \\
2018-01-03 &&  1                   &&   2458122.3461 &&  $+12.85\pm1.50$  &  9.6 &&  $-18.89\pm1.50$  & 10.5 && $+22.02\pm2.00$ & 10.0 && $ -7.52\pm2.00$ & 10.0 && FIES    \\
2018-11-23 &&  1                   &&   2458446.4436 &&  $-15.08\pm1.50$  & 10.0 &&  $+12.90\pm1.50$  & 12.5 && $ -3.00\pm2.00$ & 10.1 && $+20.49\pm2.00$ & 10.0 && FIES    \\
2018-11-26 &&  1                   &&   2458449.3208 &&  $-14.90\pm1.50$  & 10.0 &&  $+14.01\pm1.50$  &  8.0 && $ -5.88\pm2.00$ & 10.0 && $+24.75\pm2.00$ &  7.8 && FIES    \\
2018-12-01 &&  1                   &&   2458454.3734 &&  $-14.82\pm1.50$  &  9.8 &&  $+13.23\pm1.50$  & 14.5 && $ -6.21\pm2.00$ & 10.2 && $+21.79\pm2.00$ &  8.7 && FIES    \\
2018-12-07 &&  1                   &&   2458460.3696 &&  $-15.12\pm2.00$  & 10.3:&&  $+12.52\pm2.00$  & 17.3:&&                 &      && $+21.66\pm3.00$ & 21.7:&& FIES    \\
2018-12-10 &&  1                   &&   2458463.4205 &&  $-15.29\pm1.50$  & 10.0 &&  $+14.43\pm1.50$  &  8.0 && $ -4.01\pm2.00$ & 10.0 && $+22.01\pm2.00$ & 10.0 && FIES    \\
2018-12-22 &&  1                   &&   2458475.3701 &&  $-15.09\pm1.50$  & 10.7 &&  $+14.35\pm1.50$  &  8.1 &&                 &      && $+22.37\pm2.00$ & 10.6 && FIES    \\
2018-12-24 &&  1                   &&   2458477.3687 &&  $-14.26\pm1.50$  & 10.7:&&                   &      &&                 &      &&                 &      && FIES    \\
2018-12-26 &&  1                   &&   2458479.3677 &&  $-14.69\pm1.50$  & 10.1 &&  $+14.39\pm1.50$  &  8.0 && $ -2.37\pm2.00$ & 10.9 && $+22.01\pm2.00$ &  9.0 && FIES    \\
2018-12-27 &&  1                   &&   2458480.3449 &&  $-15.53\pm1.50$  & 10.0 &&  $+15.27\pm1.50$  &  9.3 && $ -5.03\pm2.00$ & 10.5 && $+22.25\pm2.00$ &  8.9 && FIES    \\
2019-01-02 &&  1                   &&   2458486.3505 &&  $-14.80\pm1.50$  &  9.6 &&  $+16.06\pm1.50$  &  8.2 && $ -5.28\pm2.00$ & 10.0 && $+24.16\pm2.00$ &  8.0 && FIES    \\
2019-01-06 &&  1                   &&   2458490.3547 &&  $-14.84\pm1.50$  &  9.8 &&  $+17.04\pm1.50$  &  8.5 && $ -5.48\pm2.00$ & 12.4 && $+22.01\pm2.00$ &  8.1 && FIES    \\
2019-01-08 &&  1                   &&   2458492.3427 &&  $-14.15\pm1.50$  &  9.6 &&  $+13.04\pm1.50$  &  7.9 && $ -4.20\pm2.00$ &  9.9 && $+22.95\pm2.00$ &  8.0 && FIES    \\
\hline\noalign{\smallskip}
\multicolumn{19}{l}{Notes: `N' is the number of spectra. ($\dagger$): Radial velocities  and velocity dispersions ($b$) for these two dates have been taken from
  \cite{Welsh98}, since the observations}\\
\multicolumn{19}{l}{\hspace{0.7cm} were not available for direct measurements following the same method used in this work. (:) Noisy spectra.}
\end{tabular}
\end{table*} 


\subsection{Radial velocities and velocity dispersions.}
\label{Sect:RVs}

Table \ref{Table:RVs} shows the RVs and velocity dispersions, $b$, of
the deep, D, and shallow, S, components of the narrow CS absorption
features superimposed on the bottom of the photospheric Ca {\sc ii} K line,
measured for all spectra. In those cases where a four-Gaussian fit
was feasible, the same results for the d and s components are listed.
The RVs correspond to the minimum of the main Gaussian components of
the absorption (see Sect. \ref{Sect:CS_FEATURES}) and the dispersion
velocities are the full widths of the Gaussian profiles. All the values
correspond to spectra obtained on individual nights; only in three
cases were the measurements carried out over averaged spectra
obtained on consecutive nights, namely 28-29 Aug 2004 (CS21),
4 to 7 Sep 2015 (HERMES), and \mbox{23-24} Oct 2017 (HARPS-N), to
increase the S/N of the profile.

Uncertainties were assigned in a very conservative way taking
into account those provided by the \texttt{ELF} package after the
Gaussian decomposition of the profiles, and the effect of potential
offsets between the wavelength calibrations provided by the pipelines
of the different instruments. The quantitative effects of these offsets
can be estimated by measuring the position of the telluric lines in
spectra from the different spectrographs; differences in velocities were
found to be less than $\sim\!0.2$ km/s.

\begin{figure}
\centering
\includegraphics[width=0.5\textwidth]{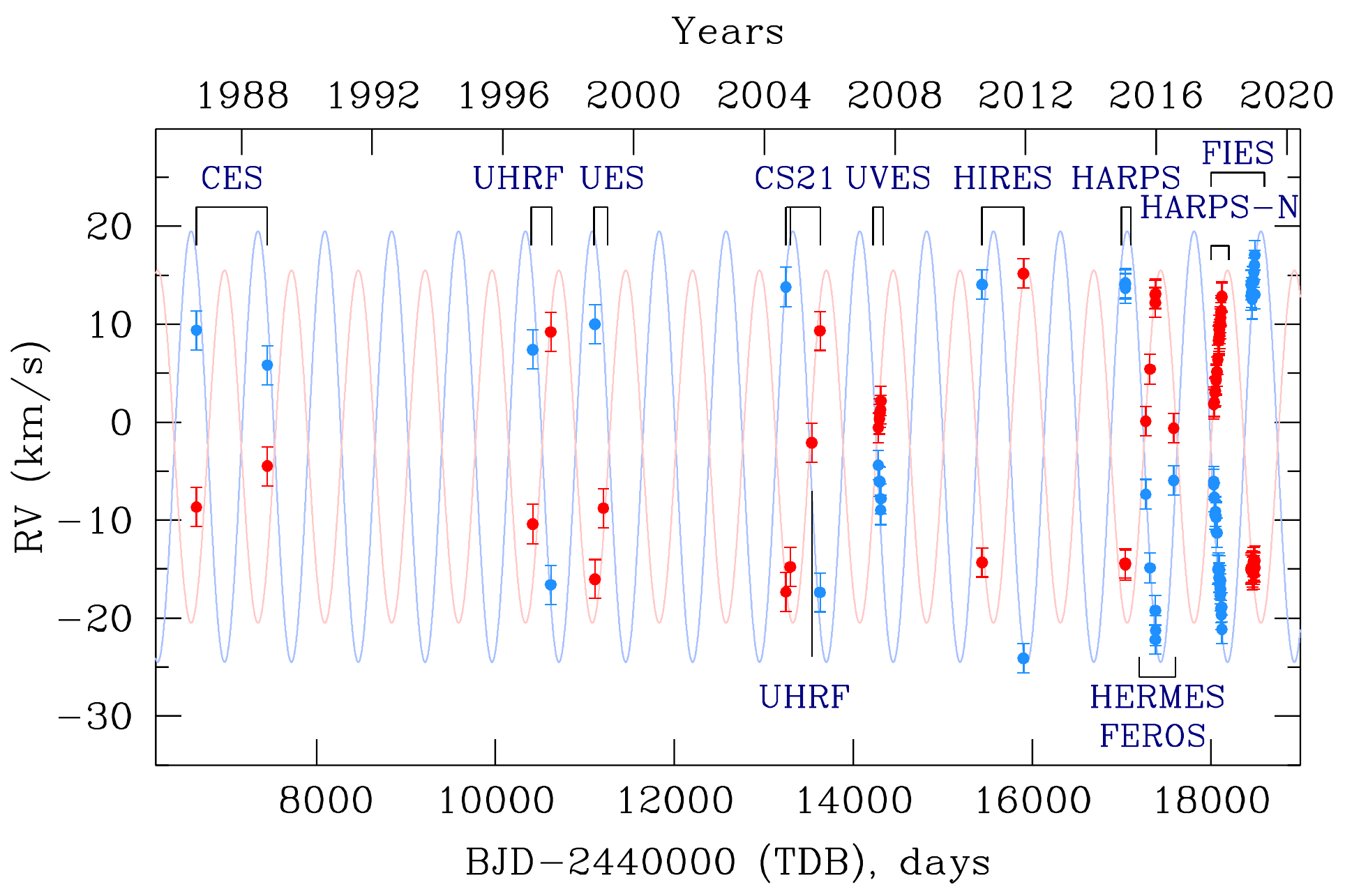}
\caption{Radial velocities of the components D (red) and S (blue)  of
  the CS Ca {\sc ii} K narrow absorption features for all the
  observations. The sinusoidal curves plotted as pale red and
  blue have the only purpose to show, in a qualitative way, that the
  behaviour of the RVs shows a periodic pattern during the
  $\sim\!32$-yr interval covered by the observations. The exact
  treatment of the data is explained in
  Sect. \ref{Sect:Spec_binary_sol}. The labels indicate the
  instruments with which the observations were obtained (see Sect.
  \ref{Sect:SPECTRA}).}
\label{Fig:RVs}
\end{figure}

Figure \ref{Fig:RVs} shows a plot of the whole set of RVs for the D
(red) and S (blue) components given in Table \ref{Table:RVs} with the
corresponding error bars. With the only purpose of guiding the
  eye, we have included in the graph two sinusoidal curves, plotted
in pale red and blue, without any quantitative physical
  meaning, to show that the behaviour of the RV of each component
seems to follow a periodic pattern over decades. The most plausible
scenario is that each component, D and S, traces the orbit of
  each individual star in the binary. In the following sections we prove,
by means of robust calculations of a spectrometric binary solution
(Sect. \ref{Sect:Spec_binary_sol}) and an orbital solution
(Sect. \ref{Sect:Orbital_binary_sol}), that this hypothesis is
consistent.


\begin{table}
\caption{Median and 68.7\% confidence intervals for the RV parameters analysed in Sect.
\ref{Sect:Spec_binary_sol}. Prior distributions are also given.}
\label{Table:RV_res}
\begin{tabular}{lll}
\hline \hline
\noalign{\smallskip}
Parameter & Prior & Posterior \\
\noalign{\smallskip}
\hline
\noalign{\smallskip}
$V_{\rm sys}$ [km/s]         & $\mathcal{U}(-30.0,30.0)$      & $-2.18^{+0.34}_{-0.31}$  \\
\noalign{\smallskip}
$P_{\rm orb}$ [days]         & $\mathcal{G}(747.0,20.0)$      & $747.60^{+0.61}_{-0.70}$ \\
\noalign{\smallskip}
$T_{\rm c,A}$ [BJD-2400000]   & $\mathcal{G}(57562.4,20.0)$    & $57563.5^{+8.0}_{-7.2}$  \\
\noalign{\smallskip}
$K_{\rm A}$ [km/s]           & $\mathcal{U}(0.0,40.0)$        & $16.45^{+0.76}_{-0.79}$  \\
\noalign{\smallskip}
$e_{\rm A}$                  & $\mathcal{U}(0.0,0.9)$         & $0.254^{+0.036}_{-0.039}$ \\
\noalign{\smallskip}
$\omega_{\rm A}$ [$^{\circ}$] & $\mathcal{U}(-180.0,180.0)$    & $38.0^{+4.0}_{-4.0}$     \\
\noalign{\smallskip}
$T_{\rm c,B}$ [BJD-2400000]   & $\mathcal{U}(57562.0,58309.0)$ & $58022.3^{+6.4}_{-8.6}$  \\
\noalign{\smallskip}
$K_{\rm B}$ [km/s]           & $\mathcal{U}(0.0,40.0)$        & $20.35^{+0.69}_{-0.72}$  \\
\noalign{\smallskip}
$e_{\rm B}$                  & $\mathcal{U}(0.0,0.9)$         & $0.209^{+0.030}_{-0.026}$\\
\noalign{\smallskip}
\hline
\noalign{\smallskip}
\multicolumn{3}{l}{Note: $\mathcal{U}(a,b)$ stands for uniform prior between $a$ and $b$, while}\\
\multicolumn{3}{l}{$\mathcal{G}(\mu,\sigma)$ represents a Gaussian prior with mean $\mu$ and standard}\\
\multicolumn{3}{l}{deviation $\sigma$.}
\end{tabular}
\end{table}


\subsection{Spectrometric binary solution}
\label{Sect:Spec_binary_sol}

In this section we derive the spectrometric solution for the binary under the
assumption mentioned in the last paragraph. We assign component D to HR 10-A,
and component S to HR 10-B, the reason for this being, as we show in the
following, that component D originates in the more massive of the two stars 
(component A) as it displays a smaller amplitude in the RV variation than
component S.

\begin{figure}
\centering
\includegraphics[width=0.5\textwidth]{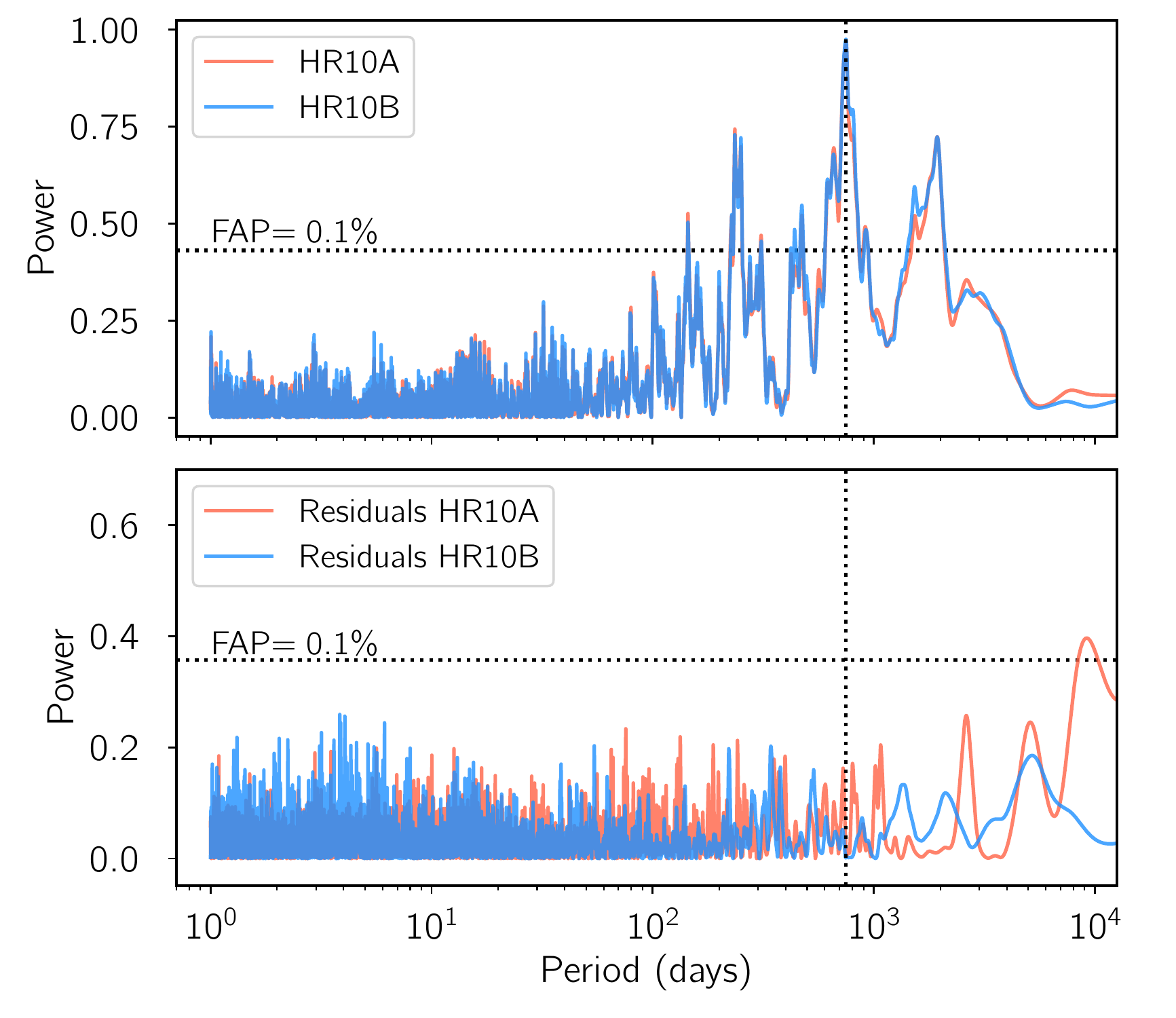}
\caption{Top: Periodogram corresponding to the RV
  measurements for each of the two components of the binary system in
  HR\,10. The 0.1\% FAP is shown as a horizontal
  dotted line and the final orbital period is displayed as a vertical
  dotted line. Bottom: Periodogram of the residuals of the
  RV modelling.}
\label{Fig:RV_periodogram}
\end{figure}

We first computed the Lomb-Scargle periodogram of the measured RVs for each
star using the \texttt{astroml} package \citep{Ivecic14} based on the methodology
described  in \cite{zechmeister09}. This periodogram is shown in the upper panel
of Fig.~\ref{Fig:RV_periodogram}. The result clearly shows a common peak above
the 0.1\% false-alarm probability (FAP) corresponding to a periodicity of
$\approx\!747$ days, indicating that both are gravitationally bound. Although the
periodogram shows other peaks above the 0.1\% FAP level, we attribute their
origin to the different offsets between the many different instruments used,
which are not taken into account in this periodogram computation but are in
the modelling of the RV.

We analyse the RV sets obtained from each component, D and
S, -- corresponding to stars A and B -- by modelling the whole system. 
The model includes two Keplerian orbits described by

\begin{equation}
\begin{split}
V_{\rm rad\,A} = & V_{\rm sys} + K_{\rm A}\left[ \cos(\nu_{\rm A}+\omega_{\rm A}) + e_{\rm A}\cos(\omega_{\rm A}) \right]  \\
V_{\rm rad\,B} = & V_{\rm sys} + K_{\rm B}\left[ \cos(\nu_{\rm B}+\omega_{\rm A}+\pi) + e_{\rm B}\cos(\omega_{\rm A}+\pi) \right],
\end{split}
\end{equation}

\noindent where $V_{\rm sys}$ is the systemic velocity of the binary
system, $\omega$ is the argument of the periastron of star A, $e_{\rm A}$
and $e_{\rm B}$ are the eccentricities of the orbits of stars A and B,
respectively, and $\nu_{\rm A}$ and $\nu_{\rm B}$ are the true anomalies of each
star, which implicitly include the times of conjunction $T_{\rm c,A}$
and $T_{\rm c,B}$. We used uninformative priors for all the parameters 
except for the orbital period, $P_{\rm orb}$, and the conjunction passage. 
For $P_{\rm orb}$, we used a normal distribution around the highest peak of the
RV periodogram.  The prior for the conjunction passage was set after
a first exploration with an uninformative prior. The boundaries of the priors for
all parameters are specified in Table \ref{Table:RV_res}. Additionally, we
included $N_{\rm inst}$-1 additional parameters to account for RV
offsets between the different instruments, with $N_{\rm inst}$ being the number 
of instruments used. We set uninformative priors for these parameters between
$-1$ and $+1$ km/s. In total, 19 parameters were explored.

\begin{figure*}
\centering
\includegraphics[width=1.0\textwidth]{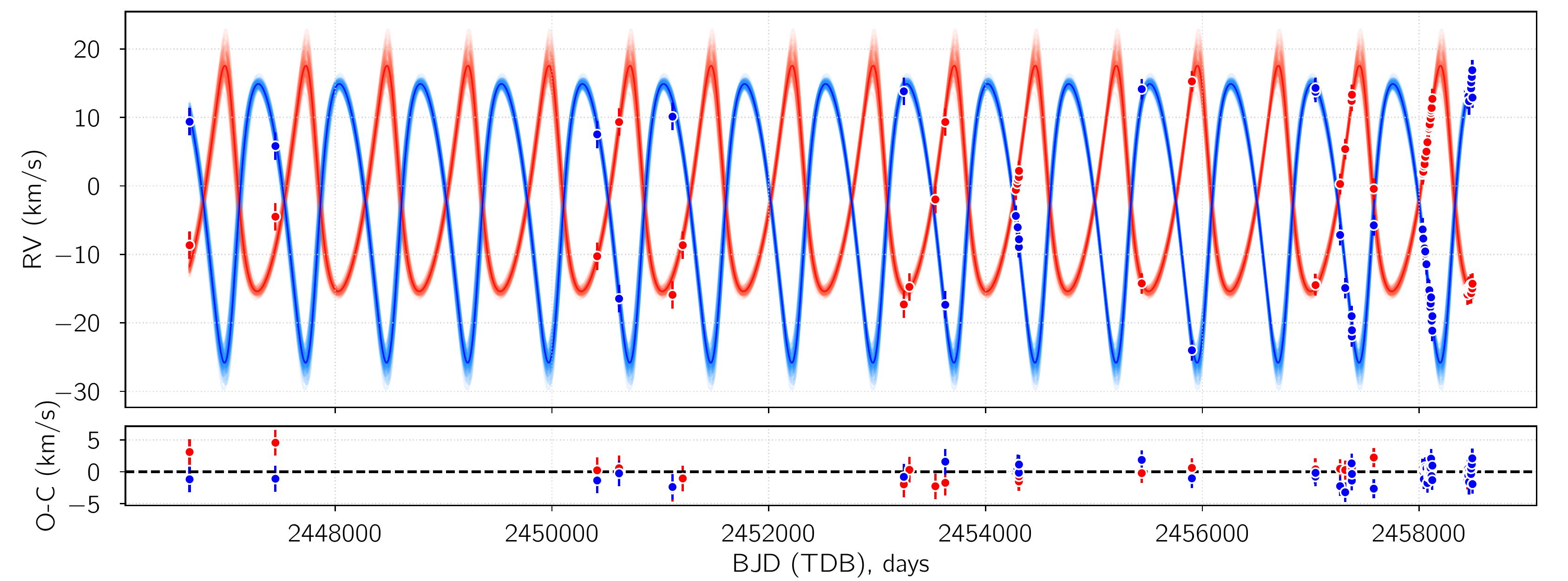}
\caption{Radial velocity measurements of the two stars along the $\sim\!32$
  years of observations. Red and blue symbols and lines correspond to 
  components D and S from the Gaussian fittings of the narrow Ca {\sc ii} K
  absorptions, which in turn correspond to stars A and B, respectively. The upper
  panel shows the model corresponding to the median values of the posterior
  distributions of the parameters as thick lines and 100 random samples drawn
  from the final MCMC chain; the lower panel shows the residuals.}
\label{Fig:RV_JD}
\end{figure*}

\begin{figure}
\centering
\includegraphics[width=0.5\textwidth]{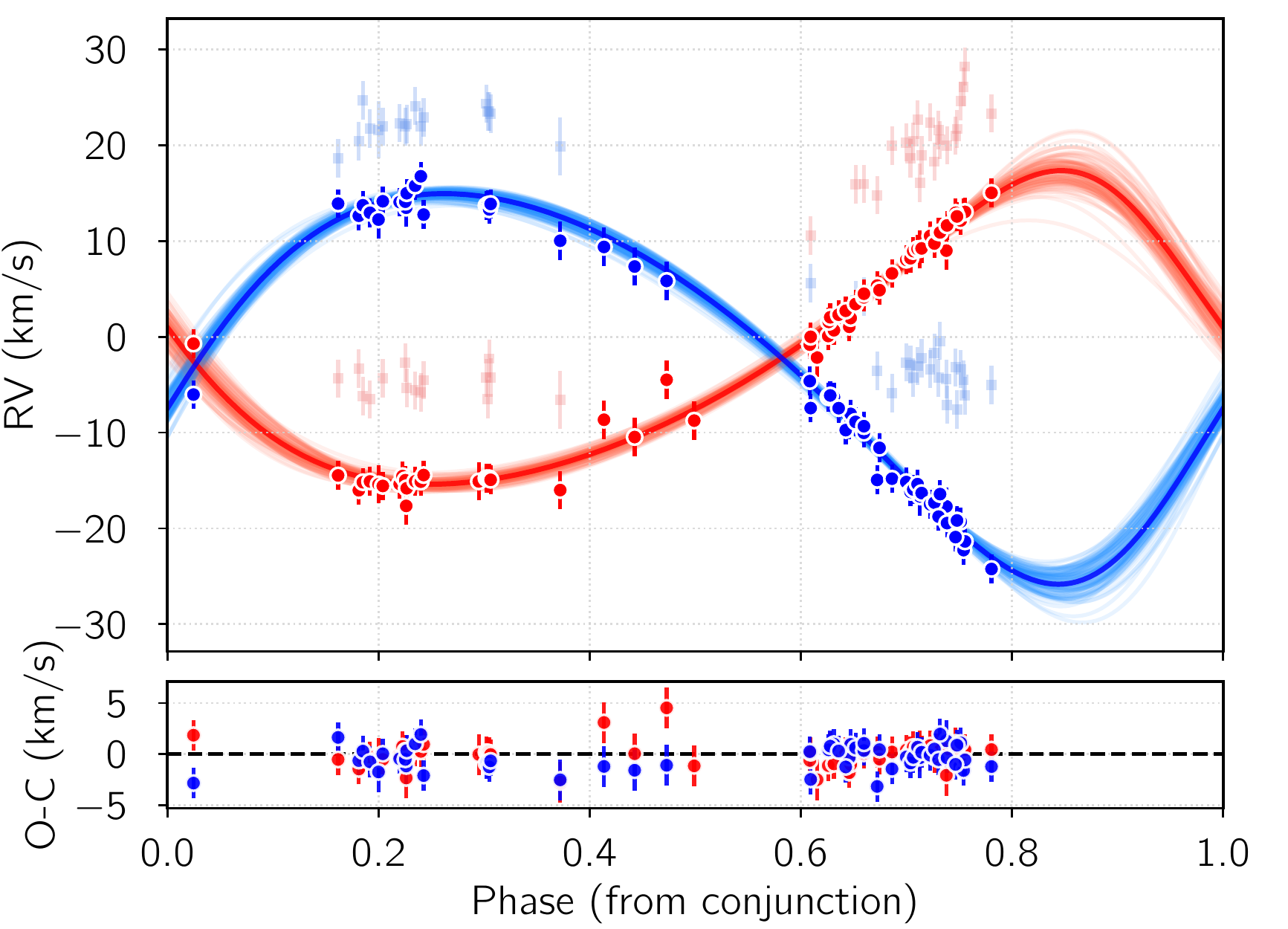}
\caption{Phase-folded RV curves of the two stellar
  components of HR\,10. Red and blue circles and lines follow the same
  colour code as in Fig.~\ref{Fig:RV_JD}. Pale red and blue squares correspond 
  to the RVs of components d and s; further details of 
  their behaviour are given in Sect. \ref{Sect:DSds}.}
\label{Fig:RV_phase}
\end{figure}

In order to sample the posterior probability distribution of each of
those parameters, we used the implementation of Goodman \& Weare's
affine invariant Markov chain Monte Carlo (MCMC) ensemble sampler
\texttt{emcee} developed by \cite{Foreman-Mackey13}. We used 50
walkers and 5000 steps per walker. This turned out to be enough due to
the quick convergence of the chains. In order to compute the final
posterior distributions, we discarded the first half of the chains and
combined the second half to build the marginalised posteriors, finally
composed of $1.25\times10^5$ steps. The median and 68.7\% confidence
intervals for each parameter are shown in Table~\ref{Table:RV_res}. The
results of the final fitting are displayed in Fig.~\ref{Fig:RV_JD}
showing the whole time span and Fig.~\ref{Fig:RV_phase} showing the
phase-folded RV curves; this figure also includes the 
RVs of  components d and s,  the behaviour of which is described in
Sect. \ref{Sect:DSds}.

\begin{figure}
\centering
\includegraphics[width=0.5\textwidth]{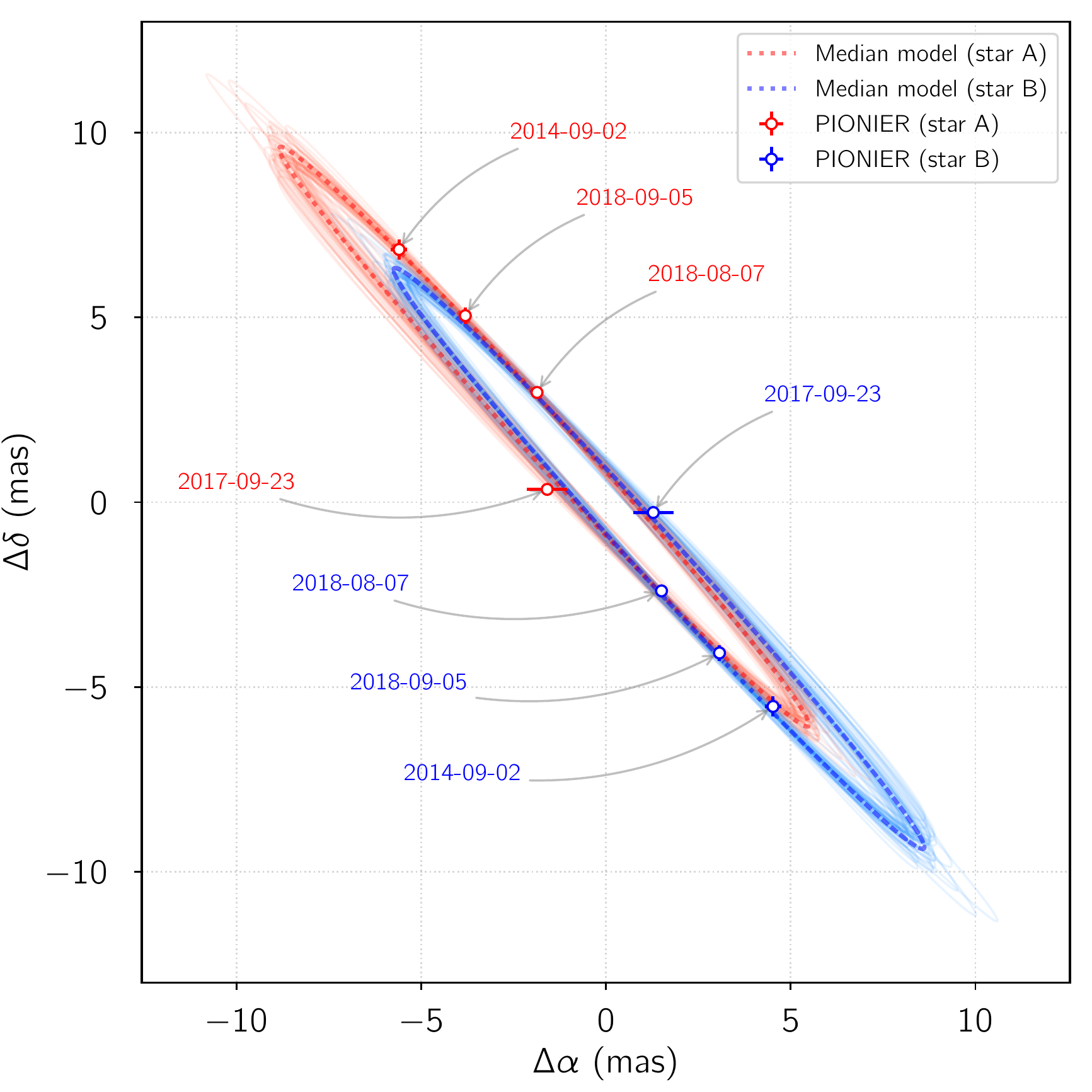}
\caption{PIONIER/VLTI astrometric positions for the two components of the
  HR10 resolved binary system (circle symbols include error bars). The
  date for each observation is specified. A sample of 100
  orbital solutions obtained from the posterior distributions derived
  in Sect~\ref{Sect:Orbital_binary_sol} (see Table
  \ref{Table:ORBITAL_res}), are represented by the colour-coded thin
  lines. The median models are represented by thick dashed lines. }
\label{Fig:ORBITS}
\end{figure}

\subsection{Orbital solution}
\label{Sect:Orbital_binary_sol}
\begin{table}
\caption{Median and 68.7\% confidence intervals for the orbital
  parameters analysed in Sect. \ref{Sect:Orbital_binary_sol}. Prior
  distributions are also displayed}
\label{Table:ORBITAL_res} 
\begin{tabular}{lll}
\hline \hline
\noalign{\smallskip}
Parameter & Prior & Posterior \\
\noalign{\smallskip}
\hline
\noalign{\smallskip}
$P_{\rm orb}$ [days]  & $\mathcal{G}(747.58441074,0.67)$ & $747.36^{+0.67}_{-0.64}$ \\
\noalign{\smallskip}
$T_0$ [BJD-2400000] & $\mathcal{U}(57200.0,57600.0)$ & $57453^{+22}_{-44}$ \\
\noalign{\smallskip}
$a_{\rm tot}$ [mas] & $\mathcal{U}(10.0,50.0)$ & $10.58^{+0.60}_{-0.37}$ \\
\noalign{\smallskip}
$\Omega_{\rm A}$  [$^{\circ}$] & $\mathcal{U}(0,360)$ & $137.65^{+0.83}_{-0.84}$ \\
\noalign{\smallskip}
$\omega_{\rm A}$ [$^{\circ}$] & $\mathcal{U}(-180.0,180.0)$ & $17^{+11}_{-22}$ \\
\noalign{\smallskip}
$i$ [$^{\circ}$]  & $\mathcal{U}(90.0,100.0)$ & $93.34^{+0.60}_{-0.63}$ \\
\noalign{\smallskip}
$e_{\rm A}$        & $\mathcal{G}(0.24,0.03)$ & $0.242^{+0.017}_{-0.012}$ \\
\noalign{\smallskip}
$e_{\rm B}$        & $\mathcal{G}(0.207,0.03)$ & $0.208^{+0.015}_{-0.011}$ \\
\noalign{\smallskip}
\hline
\noalign{\smallskip}
\multicolumn{3}{l}{Note: $\mathcal{U}(a,b)$ stands for uniform prior between $a$ and $b$, while}\\
\multicolumn{3}{l}{$\mathcal{G}(\mu,\sigma)$ represents a Gaussian prior with mean $\mu$ and standard}\\
\multicolumn{3}{l}{deviation $\sigma$.}
\end{tabular}
\end{table}

The four PIONIER astrometric positions of the two stellar components
of the binary system were analysed following the equations of motion
described in \citet[][see their Eqs. 7-15]{Mede17}. Given
the low number of epochs, we assumed Gaussian priors on the orbital
period, $P_{\rm orb}$, and the eccentricities of the two orbits, $e_{\rm A}$ and
$e_{\rm B}$, as obtained from the RV analysis. For the rest of
the parameters (namely the argument of the periastron of star A,
$\omega_{\rm A}$, the longitude of the ascending node of star A, $\Omega_{\rm
A}$, orbital inclination, $i$, time of periapsis $T_0$, and the total
semi-major axis, $a_{\rm tot}\!=\!a_{\rm A}+a_{\rm B}$, we used uninformative 
priors as stated in Table~\ref{Table:ORBITAL_res}. The posterior distribution of
these parameters was explored using again the \texttt{emcee} code. We
used 20 walkers and 10\,000 steps per walker to widely explore the
parameter space and then ran a second exploration in a smaller
parameter space centred on the maximum likelihood set of parameters
as obtained from the first phase. In this second run, we used 20
walkers and 5\,000 steps. The first 20\% of the steps for each walker
were removed and the posterior distributions were built from a total
of 80\,000 steps. All chains converged nicely into the values
presented in Table~\ref{Table:ORBITAL_res}, including the median and
68.7\% confidence intervals. As an example, the result for the argument
of the periastron obtained from this analysis is in perfect agreement
(within 2$\sigma$) with the result from the RV analysis
(see Table \ref{Table:RV_res}). Figure \ref{Fig:ORBITS} shows the positions 
of the stars in the plane of the sky  according to the four 
PIONIER/VLTI observations, and the solutions for the projected orbits 
of HR 10-A (red) and HR 10-B (blue) around the centre of mass. 

\subsection{Stellar parameters}
\label{Sect:INDIVIDUAL}

In order to compute a consistent set of stellar parameters, the observed
photometry, spectra, and some of the quantities extracted from the binary
solution, must be matched by the combination of the individual
synthetic photometry and spectra of each component. The following 
inputs have been used:

\begin{enumerate}
  
\item High-resolution optical spectra.
  
\item Low-resolution ultraviolet spectra: HR 10 was observed with the
  International Ultraviolet Explorer  \citep[{\em IUE},][]{Kondo87}. A
  merged spectrum covering the interval 115.0--332.0 nm was built
  from spectra SWP38943 and LWP18015 extracted from the {\em IUE} Newly 
  Extracted Spectra (INES) database\footnote{\url{http://sdc.cab.inta-csic.es/ines/}}.
  
\item Photometry: Optical and NIR photometry of HR 10 collected
  from several sources is given in Table \ref{Table:PHOTOMETRY}.
  Johnson $BV$ are from the SIMBAD database, and PAN-STARSS DR1 $grizy$
  and 2MASS $JHK_{\rm s}$ are taken from VizieR catalogues
  \texttt{II/349/ps1} \citep{Chambers16} and \texttt{II/246}
  \citep{Cutri03}, respectively. The calibration from magnitudes into
  fluxes was carried out using the zero points from
  \citet{Bessell79} and \citet{Cohen03} for $BV$ and $JHK_{\rm s}$,
  respectively; $grizy$ are AB magnitudes and a value of
  $F_0\!=\!3631.0$ Jy has been adopted.  The effective wavelength,
  $\lambda_{\rm eff}$, for each filter is provided.  An uncertainty of
  0.02 mag was assigned to the optical magnitudes, since no value
  is given in the corresponding catalogues.

\item Relative brightness of the stars constrained by the contrast in
  $H$-band ($F_H({\rm B})/F_H({\rm A})\simeq\!0.32$) from the PIONIER
  observations (see Table \ref{Table:PIONIER}).
  
\item Masses derived from the binary solution: Provided the inclination
  of the system is known, the individual masses of the components
  can be estimated with the following expression:
  
\begin{equation}
  M_{\rm A(B)}\,\sin^3 i\,=\,\frac{P_{\rm orb}}{2\pi G}\;(1-e_{\rm A(B)}^2)^{3/2}\,(K_{\rm A}+K_{\rm B})^2\,K_{\rm B(A)}
\label{Eq:MASSES}
.\end{equation}

Using the data in Tables \ref{Table:RV_res} and \ref{Table:ORBITAL_res} we obtain
$M_{\rm A}\!=\!1.94\pm0.15$, and $M_{\rm B}\!=\!1.62\pm0.13$ ${\rm M}_\odot$.
\end{enumerate}

\begin{table}
\setlength{\tabcolsep}{10pt}  
\caption{HR 10: Optical and NIÉRR photometry.}
\label{Table:PHOTOMETRY}
\begin{tabular}{lrl}
\hline \hline
\noalign{\smallskip}
Magnitude &$\lambda_{\rm eff}$ [nm] & \multicolumn{1}{c}{Filter} \\
\noalign{\smallskip}
\hline
\noalign{\smallskip}
6.33            &   440.0  &   $B$ Johnson/Bessell \\
6.25            &   477.6  &   PAN-STARRS/PS1.$g$  \\
6.230           &   550.0  &   $V$ Johnson/Bessell \\
6.328           &   612.9  &   PAN-STARRS/PS1.$r$  \\
6.434           &   748.5  &   PAN-STARRS/PS1.$i$  \\
6.507           &   865.8  &   PAN-STARRS/PS1.$z$  \\
6.539           &   960.3  &   PAN-STARRS/PS1.$y$  \\
$5.858\pm0.019$ &  1235.0  &   2MASS $J$           \\
$5.831\pm0.033$ &  1662.0  &   2MASS $H$           \\
$5.747\pm0.027$ &  2159.0  &   2MASS $K_{\rm s}$    \\
\noalign{\smallskip}\hline
\end{tabular}
\end{table}

The starting point of the estimation of the stellar parameters is a
one-model fitting of the optical high-resolution spectrum, using
Kurucz synthetic models; the method is described in
\citet{Rebollido18}; the programmes {\sc synthe} and {\sc atlas}
\citep{Kurucz93} fed with models describing the stratification of the
stellar atmospheres \citep{Castelli03} have been used for the spectral
synthesis. Since the computation of the effective temperature hinges
on the match of the depth of the photospheric Ca {\sc ii} K profile,
we derived a model that slightly underestimates that depth,
measured on a spline fitted to the line profile, avoiding the CS
components. The reason for this is that the final fit will be composed
of two models, therefore the addition of the synthetic Ca {\sc ii} K
profiles of both stars should match the observed profile. The spectral
fit was made on observations obtained with the stars almost in
conjunction, to avoid large shifts between the spectral lines of both
components.  After some iterations, the result is $T_{\rm
  eff}\!=\!9000\pm100$ K, $\log g_*\!=\!3.8\pm0.1$; the model was
broadened with $v \sin i\!=\!294$ km/s \citep{Mora01}.  We make the
initial hypothesis that the parameters of star A are close to these
ones; this hypothesis is tested at the end of the fitting
process.

Since the contrast in $H$-band between the two components is 0.32, it
is trivial to deduce that $\Delta H({\rm A}\!-\!{\rm B})\!=\!-1.24$.
Magnitudes $B_{\rm A}$, $V_{\rm A}$, $J_{\rm A}$, $H_{\rm A}$ and
$K_{\rm s, A}$ for star A, with $T_{\rm eff}$(A)=9000 K and $\log
g_*$(A)=3.8 have been computed. In order to estimate the parameters of
star B, a grid of magnitudes $B_{\rm B}$, $V_{\rm B}$, $J_{\rm B}$,
$H_{\rm B}$ and $K_{\rm s, B}$ was created, covering a range of
$T_{\rm eff}$ from 7500 to 9000 K (step 50 K), and $\log g_*$ from 3.8
to 4.3 (step 0.1 dex).  The empirical colour--temperature calibration
by \citet{Worthey11} and the code \texttt{mash3.f} provided by these
authors were used\footnote{\url{http://astro.wsu.edu/models/}}.
The sets of magnitudes were computed from the corresponding
colours keeping $\Delta H({\rm A}\!-\!{\rm B})\!=\!-1.24$.

Combined colours $B\!-\!V$, $V\!-\!H$, $J\!-\!H$ and $J\!-\!K_{\rm s}$
were computed using the magnitudes of both stars and reddened
with values of $E(B\!-\!V)$ between 0.0 and 0.2; the coefficients
$A_\lambda/A_V$ were taken from \citet[][Table 3]{Wang19}; independent
extinctions were allowed for each star in order to take into account
potential differences caused by the individual CS envelopes. These
synthetic colours are compared with the observed ones, this process
yielding a subset of solutions [9000 K, 3.8] + [$T_{\rm eff}$(B),
  $\log g_*$(B)] for which the observed and synthetic colours are in
agreement, within the uncertainties.

In order to break the degeneracies and further constrain the
parameters for star B, the observed SED, which includes the available 
photometry and the {\em IUE} spectrum, is compared with composite Kurucz
low-resolution models, built as the sum of the model for star A, 
[9000 K, 3.8], and those for [$T_{\rm eff}$(B), $\log g_*$(B)] obtained 
in the previous step. The models for stars A and B, reddened with 
the corresponding values of $E(B\!-\!V)_{\rm A}$ and $E(B\!-\!V)_{\rm B}$
are scaled so that $F_H({\rm B})/F_H({\rm A})\!=\!0.32$ at the $H$-band,
and the sum is normalized to the observed SED.

A further constraint can be obtained from the HR diagram, by imposing that
the point [$T_{\rm eff}$(B), $\log g_*$(B)] falls on the same
isochrone as the point [9000 K, 3.8]. PARSEC V2.1s evolutionary
tracks and isochrones\footnote{\url{https://people.sissa.it/~sbressan/parsec.html}}
\citep{Bressan12} with solar metallicity, $Z\!=\!0.017$, were used.

Once all these requirements are taken into account and fulfilled, a
solution for each star is found. Table \ref{Table:PARAMETERS}
shows the parameters of the individual stars, and the observed colours
compared with the synthetic ones.  The synthetic magnitudes of 
the combined pair A+B, computed by imposing $H\!=\!5.83\pm0.03$, 
are $B\!=\!6.35\pm0.06$, $V\!=\!6.22\pm0.05$, $J\!=\!5.89\pm0.05$, and 
$K_{\rm s}\!=\!5.81\pm0.06$; both these magnitudes and the synthetic 
colours are in excellent agreement with the observed photometry 
(see Table \ref{Table:PHOTOMETRY}).

Two sets of values for the masses and luminosities of the stars are
given in Table \ref{Table:PARAMETERS}. Regarding the masses, those
labelled ``from tracks'' were derived by locating the points
[$T_{\rm eff}$, $\log g_*$], [9000 K, 3.8] (A), and [8750, 4.2] (B),
which are direct results of the above analysis, on the HR diagram
$\log g_*$ -- $T_{\rm eff}$, whereas those labelled ``from eq. (\ref{Eq:MASSES})''
are computed using that equation, which contains parameters obtained
from the binary solution. The corresponding mass ratios are
$0.72\pm0.05$ and $0.84\pm0.09$, which agree with each other within the 
uncertainties; there is however a discrepancy between the two 
values of the mass for star A, the reason could be the strong dependence
of the stellar mass on $\sin^3 i$ (see eq. \ref{Eq:MASSES}).

\begin{figure}
\centering
\includegraphics[width=0.40\textwidth]{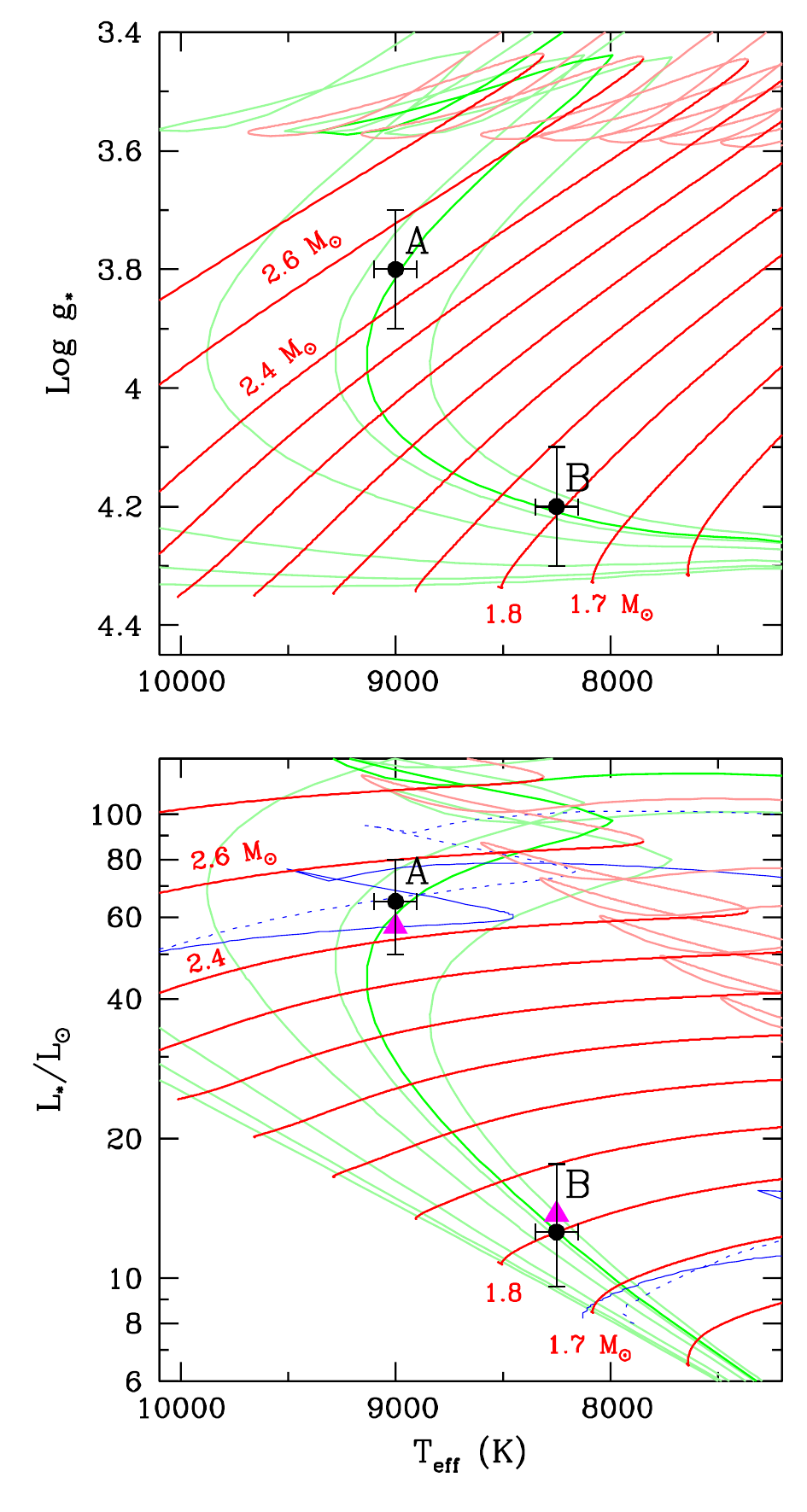}
\caption{HR diagrams $\log g_*$ -- $T_{\rm eff}$ (top) and
  $L_*/L_\odot$ (bottom) and the positions of the individual
  components of HR 10. PARSEC V2.1s evolutionary tracks (red)
  and isochrones (green) are used. Isochrones correspond to 50, 100,
  200, 400, 500, 530 (highlighted) and 600 Myr. Solid (dotted) blue
  lines in the lower panel are the evolutionary tracks with no
  rotation (rotation) for stars with 1.7 and 2.5 $M_\odot$ from
  \citet{Ekstroem12} (see Sect. \ref{Sect:CAVEATS}).  The two magenta triangles 
  in the lower HR diagram show luminosities, described below in the main text.}
\label{Fig:HR_diagrams}
\end{figure}

Regarding the luminosities, the first set of values is found by
`translating' the points [9000 K, 3.8] (A) and [8750 K, 4.2] (B) into
the tracks in the $L_*/L_\odot$ -- $T_{\rm eff}$ HR diagram.
Figure \ref{Fig:HR_diagrams} shows the HR diagrams $\log g_*$ (top) and
$L_*/L_\odot$ (bottom) -- $T_{\rm eff}$ with the positions of the
individual components of HR 10, plotted as black dots with the corresponding
error bars. The PARSEC V2.1s evolutionary tracks and isochrones are plotted 
in red and green, respectively. The stretch of the tracks plotted in a 
more intense red tone correspond to the time span where hydrogen is 
actively burning in the core.

The estimate of the age of the system is derived from the position of
the stars in the HR diagram. Below, we give details about the second
set of values for the luminosities in Table \ref{Table:PARAMETERS}, and 
those shown as magenta triangles in the bottom panel of Fig. \ref{Fig:HR_diagrams}.

\begin{table}
\setlength{\tabcolsep}{8pt}  
\caption{HR 10: Stellar parameters and colours.}
\label{Table:PARAMETERS}
\begin{tabular}{lcc}
\multicolumn{3}{l}{Parameters of the individual components}\\
\hline \hline
\noalign{\smallskip}
Parameter & Star A & Star B \\
\noalign{\smallskip}
\hline
\noalign{\smallskip}
$T_{\rm eff}$ [K]                          &  $9000\pm100$   &  $8250\pm100$  \\
$\log g_*$ [cgs]                           &   $3.8\pm0.1$   &  $4.2\pm0.1$   \\
$L/L_\odot$ [from tracks]                  &  $64.9\pm10.0$  & $12.6\pm4.0$   \\
$L/L_\odot$ [from SED fit]                 &  $57.3\pm2.0$   & $13.7\pm0.5$   \\
$M/M_\odot$ [from tracks]                  &  $2.5\pm0.1$    &  $1.8\pm0.1$   \\
$M/M_\odot$ [from eq. (\ref{Eq:MASSES})]  &  $1.94\pm0.15$  &  $1.62\pm0.13$ \\
$v \sin i$ [km/s]                          &  $294\pm9$      &  $200\pm20$    \\
Age [Myr]                                  & \multicolumn{2}{c}{$530\pm50$}   \\
$E(B\!-\!V)$                               &  0.10           &    0.05        \\
\noalign{\smallskip}\hline
\noalign{\bigskip}
\multicolumn{3}{l}{Observed and synthetic colours}\\
\hline\hline\noalign{\smallskip}
Colour           &    Observed      &    Synthetic (A+B) \\
\noalign{\smallskip}
\hline
\noalign{\smallskip}
$B\!-\!V$        & $+0.10\pm0.04$   & $+0.13\pm0.04$ \\
$V\!-\!H$        & $+0.40\pm0.04$   & $+0.40\pm0.04$ \\
$J\!-\!H$        & $+0.03\pm0.04$   & $+0.06\pm0.04$ \\
$J\!-\!K_{\rm s}$  & $+0.11\pm0.04$  & $+0.08\pm0.04$\\
\noalign{\smallskip}\hline
\end{tabular}
\end{table}

\begin{figure}
\centering
\includegraphics[width=0.5\textwidth]{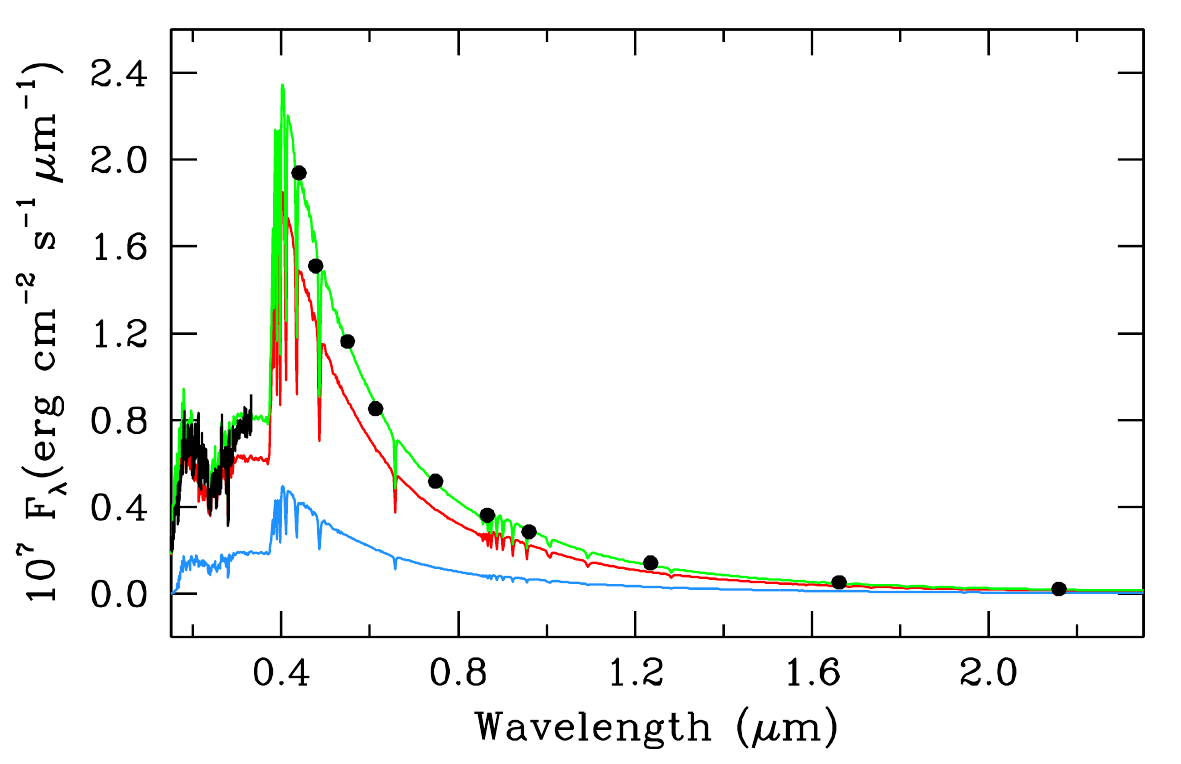}
\caption{Spectral energy distribution (SED) of HR 10 and the adopted
  solution. The observed optical and NIR photometry and the ultraviolet
  {\em IUE} spectrum SWP38943 + LWP18015 are plotted in black. The models for
  stars A and B and the combined solution are plotted in red, blue, and green,
  respectively.} 
\label{Fig:SED_fit}
\end{figure}

\begin{figure*}[t]
\centering
\includegraphics[width=17cm]{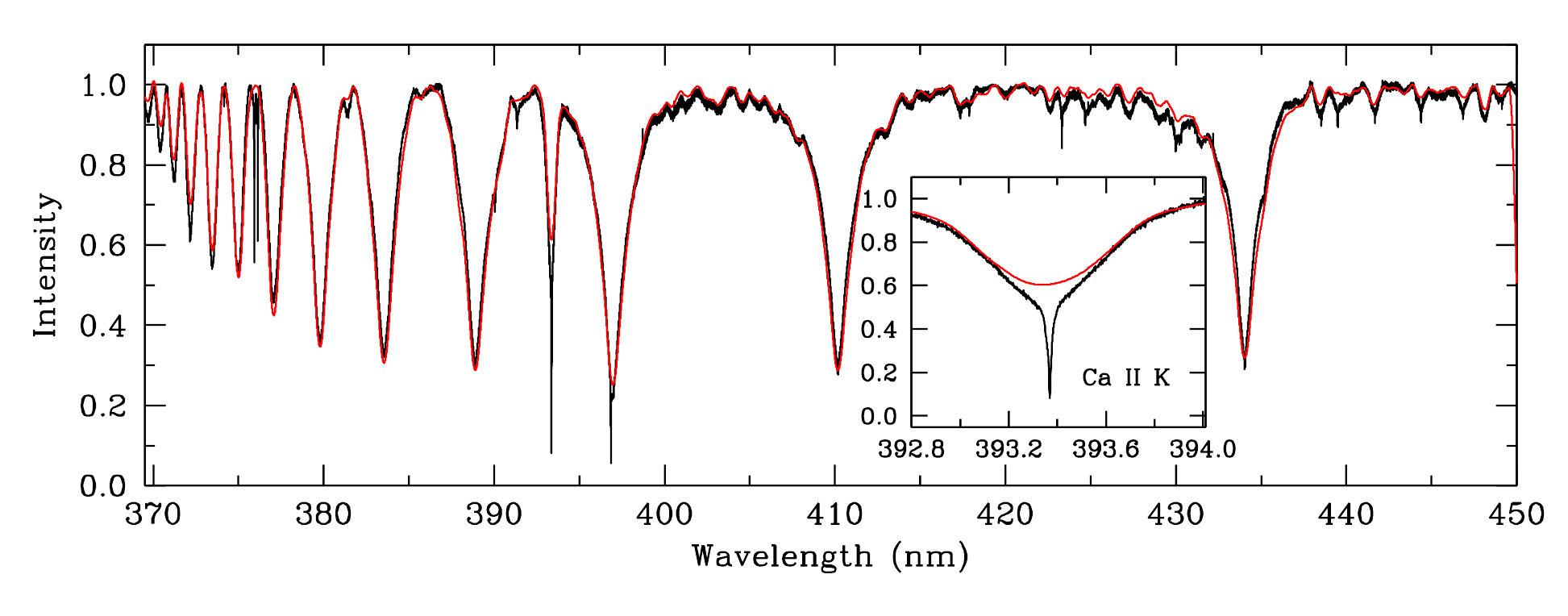}
\caption{The spectrum of HR 10 (black) and the spectral fit (red) computed
  for the composite model for the stellar parameters given
  in Table \ref{Table:PARAMETERS}. The inset shows a blow-up of the
  Ca {\sc ii} K profile and the fit. The main narrow features
  that are not reproduced by the synthetic model -- which only accounts for the
  photospheric spectrum -- are the Ti {\sc ii} lines at 375.93 and
  376.13 nm, and the Ca {\sc ii} HK lines at 393.37 and 396.85 nm (see
  Sect. \ref{Sect:CS_FEATURES}).}
\label{Fig:SPECTRAL_fit}
\end{figure*}

Figure \ref{Fig:SED_fit} shows the SED of HR 10 composed of the
ultraviolet {\em IUE} spectrum and the photometry converted into
fluxes together with the composite model solution A+B (green) and the individual 
Kurucz models for stars A (red) and B (blue) reddened with values of $E(B\!-\!V)$ 
0.10 and 0.05, and scaled so that $F_H({\rm B})/F_H({\rm A})\!=\!0.32$. 
The plot shows the contribution of each star to the total SED
very clearly. Individual luminosities for each component have been computed by
dereddening each model, integrating -- which provides values of $F_{\rm
  obs}\,({\rm erg}\,{\rm cm}^{-2}\,{\rm s}^{-1})$ at Earth -- and
converting the results into luminosities using the expression
$L_*\!=\!4\,\pi\,F_{\rm obs}\,d^2$ where $d\!=\!145.18^{+2.54}_{-2.45}$ pc 
({\em Gaia} DR2). These luminosities are shown as two magenta triangles at 
the bottom HR diagram of Fig. \ref{Fig:HR_diagrams}. The agreement between 
the values of the two sets of luminosities is excellent.

Figure \ref{Fig:SPECTRAL_fit} shows the comparison of the observed
high-resolution spectrum of HR 10 with the composite synthetic
model. The spectrum is the median of 14 UVES/VLT spectra obtained on 2007-06-28, 
and three HARPS-N spectra obtained on 2017-10-06 and 2017-10-23/24. 
These spectra were selected because the radial velocities of the stars are small 
(see Table \ref{Table:RVs}), and therefore we can consider that the stars are almost
in conjunction, and no substantial broadenings of the lines due to
large differences between the RVs of the stars are present. Model for
star A [9000 K, 3.8] was rotationally broadened with $v \sin
i\!=\!294$ km/s \citep{Mora01}. The projected rotational velocity of
200 km/s used for star B has been assigned after some iterations and
comparison with some observed photospheric features, although this
value has to be taken with caution because there is no formal way to
obtain a more robust result. The depth and full width at half depth of
the photospheric Ca {\sc ii} K profile, measured on a spline traced on
the observed line are $0.40\pm0.02$ and 484 km/s, respectively,
whereas the values in the composite A+B model are $0.39\pm0.02$ and
465 km/s. It is interesting to point out that the bottom of the Balmer
lines shows a profile that cannot be reproduced by the
synthetic spectrum, that yields a slightly more rounded shape as
a consequence of the high values of the projected rotational velocities 
of the stars. 

Given the excellent agreement between the observed quantities and
those extracted during the above thorough analysis, we can consider
that the initial hypothesis made about the parameters $T_{\rm eff}$
and $\log g_*$ for star A being close to the first estimate is 
consistent.

\section{Discussion}
\label{Sect:DISCUSSION}

\subsection{Interpretation of the D, S, d, and s profiles}
\label{Sect:DSds}

\begin{figure}
\centering
\includegraphics[width=0.45\textwidth]{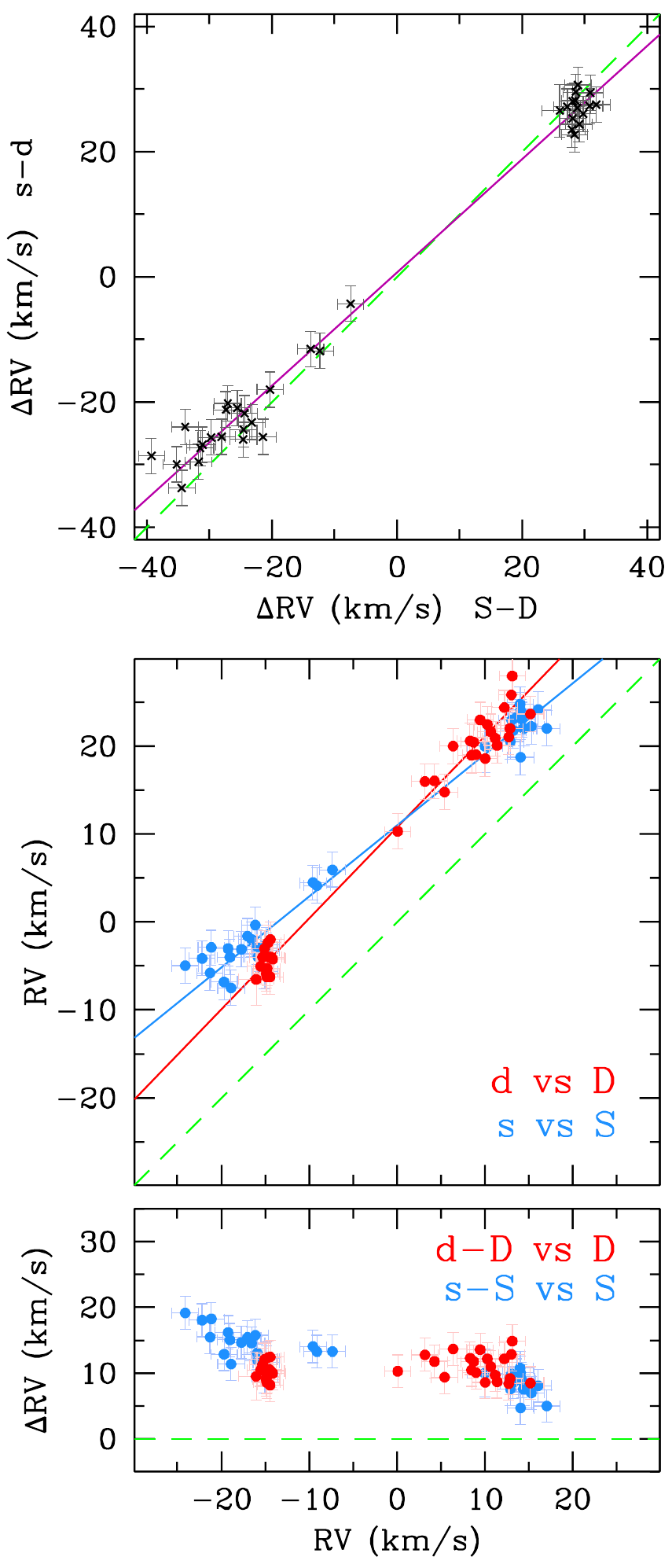}
\caption{Top: Difference between the RVs of 
  components s and d for each observation, s\,--\,d, plotted against the
  difference between the corresponding RV of components S and D,
  S\,--\,D. Middle: RV of  components d and s, plotted
  against the corresponding RV of components D and S; the dashed green 
  line is the 1:1 relationship. Bottom: Differences in RVs d\,--\,D 
  and s\,--\,S plotted against D and S, respectively.}
\label{Fig:DSds}
\end{figure}

We have shown that the RVs of the D and S narrow absorption features
trace the movements of the primary and secondary components, HR 10-A
and -B. The important result that immediately comes out of this is
that  both stars hold individual CS envelopes. Should the star
have a unique, circumbinary shell, we would only observe a single
narrow absorption.

In Fig. \ref{Fig:RV_phase}, already described in
Sect. \ref{Sect:Spec_binary_sol}, we have included the RVs of  components d
and s, plotted as pale red and blue squares. As
mentioned above, these components appear redshifted with respect to D
and S, and do not seem to be transient phenomena, since they are
clearly apparent at any time when the difference in RVs between D and
S is large enough. In addition, we show (see Fig.
\ref{Fig:CS_4lines4components}) that these weak components are also
present in the CS narrow absorptions of species other than Ca {\sc
  ii}.

In what follows we carry out a quantitative analysis of 
components d and s for the narrow CS absorptions of the Ca {\sc ii} K
line. Figure \ref{Fig:DSds} shows the behaviour of the RVs of D, S, d,
and s for all the observations where a four-Gaussian decomposition was
feasible\footnote{To avoid a tedious notation, we use the
  nomenclature S\,--\,D, s\,--\,d, d\,--\,D, and s\,--\,S to express
  the differences in radial velocities RV(S)\,--\,RV(D),
  RV(s)\,--\,RV(d), RV(d)\,--\,RV(D) and RV(s)\,--\,RV(S),
  respectively.}. The issue is to ascertain whether or not there is a
physical link between the main components D and S, and the redshifted components
d and s.

The upper panel of Fig. \ref{Fig:DSds} shows s\,--\,d plotted against
S\,--\,D, that is, the difference between the radial velocities of the weak
components, compared with the same quantity for the main
ones. The diagonal of the diagram is plotted as a dashed green line
whereas a linear fit, plotted in purple, yields a slope
$0.91\!\pm\!0.02$ \citep[an orthogonal fit, giving both variables a
  symmetric treatment, was chosen,][]{Isobe90}.  This shows that, on
average, the differences in RVs between the weak components s\,--\,d
and the main components S\,--\,D behave quantitatively in the
same way.

Some differences are apparent when individually analysing the
behaviour of  components d and s with respect to D and S. 
In the middle panel of Fig. \ref{Fig:DSds} the RVs of d
(s) are plotted against the RVs of D (S) in red (blue), the slopes of
the linear fits being $1.04\!\pm\!0.02$ and $0.81\!\pm\!0.02$,
respectively. As a general feature, components d and s 
are always redshifed with respect to the main components D and S; 
however it is interesting to note that whereas for the whole range of RVs 
of component D, the distance d\,--\,D is always within a narrow range of
$10.72\!\pm\!1.72$ km/s (see lower panel of Fig. \ref{Fig:DSds}), in the
case of  component S, the mean distance s\,--\,S is $8.29\!\pm\!1.79$
for positive values of S and $14.18\!\pm\!2.47$ for negative values; the
overall trend seems to be a decrease of s\,--\,S as the RV of component S
increases.

After this quantitative analysis, the physical link between D, S,  d, and s 
seems fairly clear, but once the transient nature of the phenomenon is discarded, 
the real origin of the weak components is uncertain.

\subsection{Evolutionary status of HR 10}
\label{Sect:EVOL}

In Fig. \ref{Fig:HR_diagrams} it can be seen that the components of HR
10 lie on the stretches of the tracks starting at the beginning of the
MS and ending at the first turning point, where almost all hydrogen in
the core is totally exhausted. During that time span, the stars are
actively burning hydrogen. According to the individual PARSEC V2.1s
track models, the ratio H/He (by mass) is $\sim\!0.34$ for HR 10-A and
$\sim\!1.36$ for HR 10-B; the initial ratio being $\sim\!2.51$.  The
stars lie on the $\sim\!530$-Myr isochrone, whereas the turning points
occur at $\sim630$ Myr and $\sim1.6$ Gyr for A and B, respectively,
away in both cases of the red giant branch phase, especially for star
B. Taking all this into account, HR 10-A would be in an evolutionary
stage close to subgiant, where HR 10-B is still very close to the
MS. Since a star remains on the main sequence as long as there is
hydrogen in the core that it can fuse into helium, we can consider
that HR 10 is still in the MS, although speaking more precisely, it
would be a very early post-MS binary.

\begin{figure}
\centering
\includegraphics[width=0.4\textwidth]{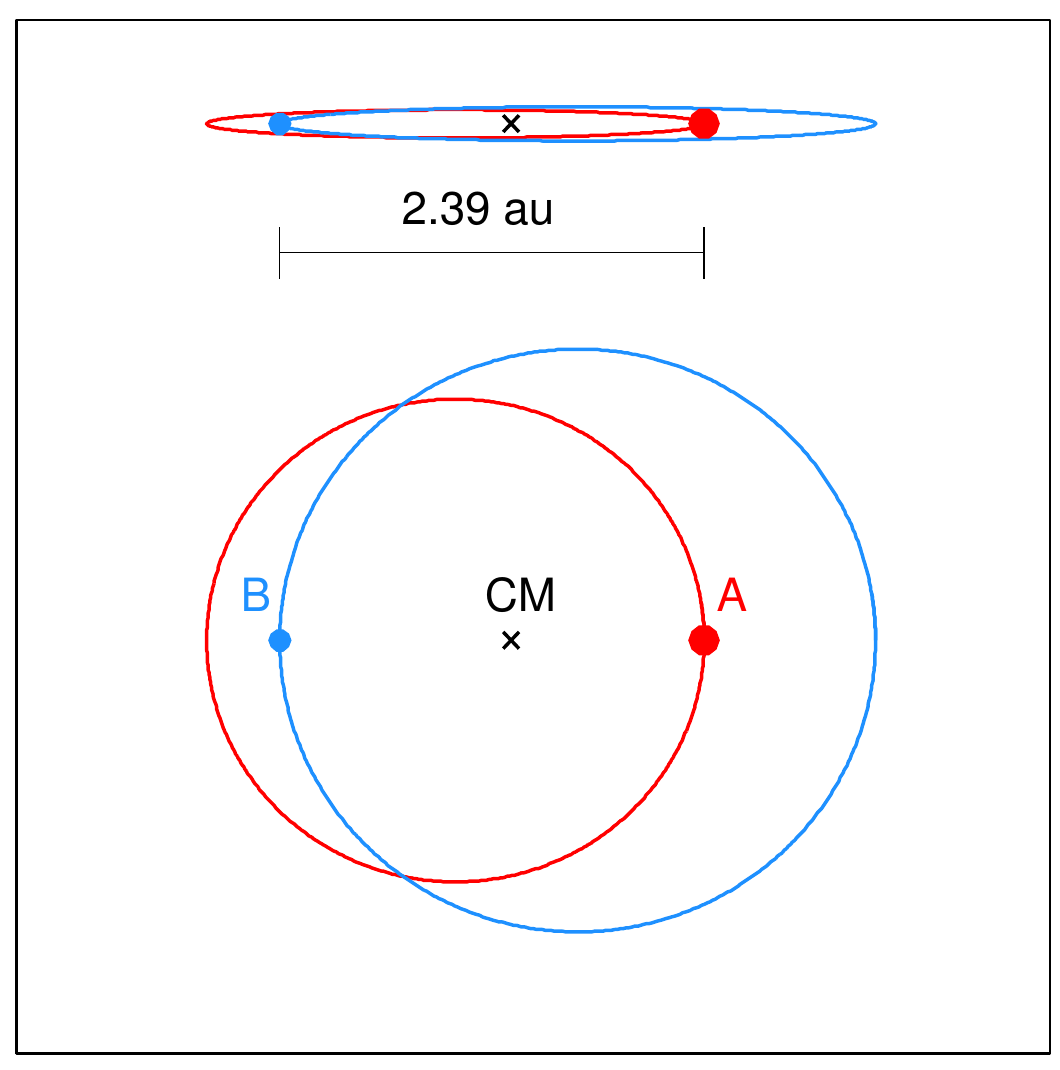}
\caption{Simple sketch of the orbits of HR 10-A and -B around the
  centre of mass of the system plotted at scale as they are
  observed (top) and deprojected onto a plane using the value of the
  inclination, $i$, from Table \ref{Table:ORBITAL_res}. The stars have
  been located at the corresponding pericentres, i.e. the points where
  the distance between components A and B is a minimum ($\sim\!2.39$
  au).}
\label{Fig:ORBIT_PLANE}
\end{figure}

\subsection{Binarity and shell/discs}
\label{Sect:BINARITY_ENVELOPES}

The value of $a_{\rm tot}$ (10.58 mas, Table \ref{Table:ORBITAL_res})
at the distance to HR 10 (145.18 pc) is equivalent to $\sim\!3.08$ au,
which implies, according to the value of the masses and
eccentricities, a minimum separation between the stars -- when both
components are at the corresponding pericentres -- of $\sim\!2.39$
au. The simple sketch of Fig.  \ref{Fig:ORBIT_PLANE} illustrates that
particular configuration, showing the orbits of HR 10-A and -B around
the centre of mass as they are observed and seen pole-on,
deprojecting the orbits using the inclination, $i$ (93.34$^\circ$,
Table \ref{Table:ORBITAL_res}). That minimum separation is a small
value, and leaving aside a dynamical and evolutionary study of the
binary, which is out of the scope of this paper, it is remarkable to
point out that i) both stars maintain their individual
envelopes or shells, and ii) the absence of a stable narrow absorption
component at the systemic velocity of the system implies that no
circumbinary envelope is present. Some studies on binarity and discs
have been carried out in other contexts; for example, Herbig AeBe stars \citep[see
  e.g.][]{Duchene15}, T Tauri stars \citep{Harris12}, and debris discs
around solar-type and intermediate-mass stars \citep{Rodriguez12}.
However, to the best of our knowledge, there have been no studies devoted to
the MS or early post-MS binary systems with shells. Therefore, this system 
is the first of its class to be discovered and studied.

\subsection{TESS photometry: rotation and pulsations}

\begin{figure*}
\centering
\includegraphics[width=1.\textwidth]{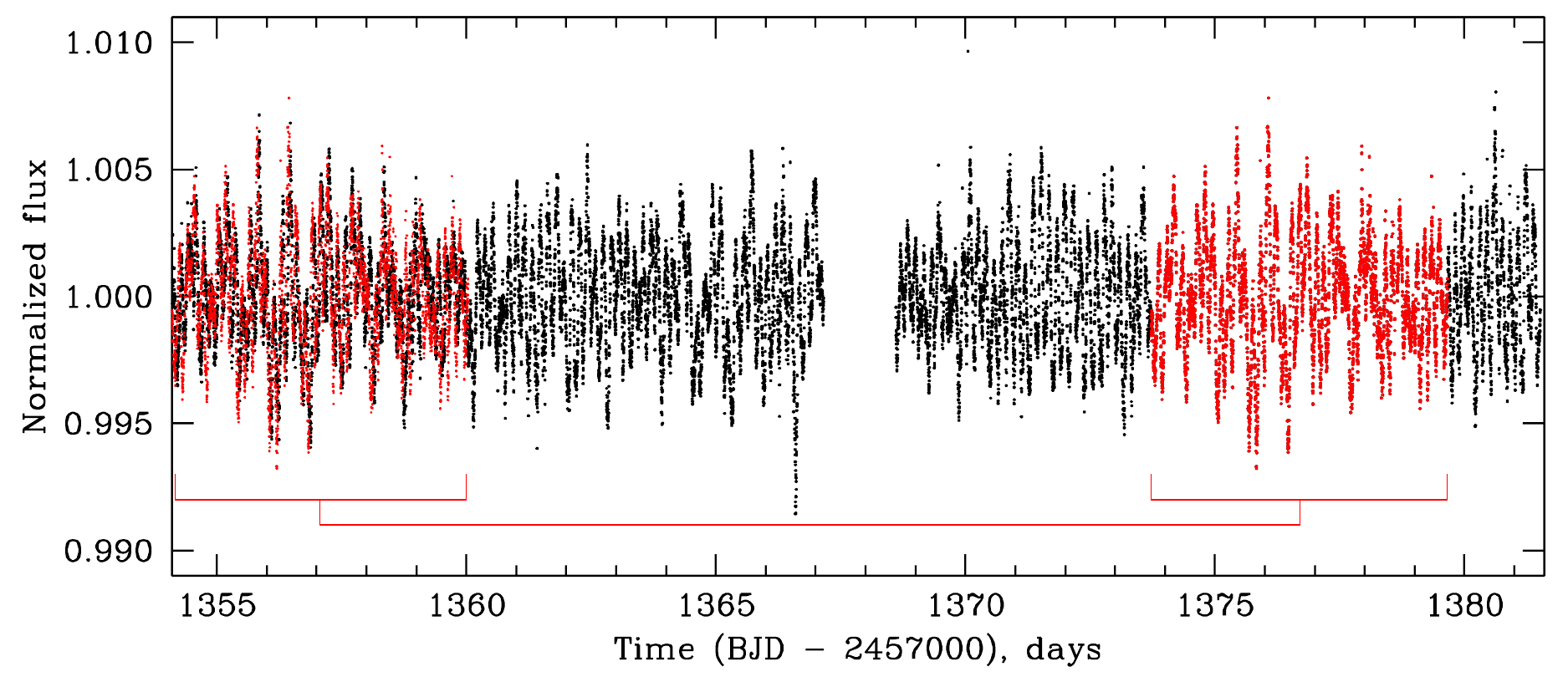}
\caption{{\em TESS} photometric time series, showing the very short-time
  variability of the HR 10 system. As an example of the repeatibility 
  of the pattern, the observations between days 1373.72 and 1379.66 
  (in BJD--2457000) highlighted in red have been shifted to the interval 
  1354.11 -- 1360.05, where they are overplotted in red on the original
  set of data. See text for details.}
\label{Fig:TESSphot}
\end{figure*}

For the sake of completeness, we would like to mention the observations
of HR 10 obtained by the Transit Exoplanet Survey Satellite mission
({\em TESS}) during Sector 2 in Camera 1 and CCD 3 (TIC\,289592423). The
photometric observations span a total of 27.4 days with a short
cadence of 2.15\,min. The data were downloaded from the Mikulski
Archive for Space Telescopes\footnote{\url{http://archive.stsci.edu}}
(MAST), including the extracted light curve by the Science Processing
Operations Center (SPOC) pipeline. The median photometric uncertainty
per datapoint for the pre-search data conditioning simple aperture
photometry (known as PDCSAP) is 140 parts per million. Given the 
large pixels and apertures used to extract the light
curve, the binary system is not resolved by {\em TESS}. In Fig.
\ref{Fig:TESSphot} the light curve shows clear periodic photometric 
variations with $\sim$5\,mmag semi-amplitude. As an example of how
the variability pattern repeats itself, the observations corresponding
to $\sim\!6$ days at the end of the observing window -- highlighted in red -- have
been shifted in time by $\sim\!-19.630$ days and superimposed on the
original data at the beginning of the observing run, fine-tuning the 
shift and causing the maxima and minima to coincide. The residuals between 
the shifted and the original data are less than $5.0\times10^{-3}$, the 
average of the absolute values being $1.0\times10^{-3}$. 
The analysis of these data, which include effects of rotation, 
pulsation, and gravitational darkening from the two components of 
HR 10, will be presented in another paper (Barcel\'o-Forteza et al.,
in preparation).

\subsection{Caveats}
\label{Sect:CAVEATS}

Finally, we would like to point out explicitly some caveats
that must be taken into account when using and interpreting the 
results presented here.

The projected rotational velocities of the stars are large, 
and therefore they must produce geometrical 
distortions which imply an oblateness of the objects, which in turn
results in a larger (smaller) gravity and temperature
at the poles (equator), the so-called gravitational darkening. Further
phenomena, such as differential rotation, prevent us from assigning a
single temperature, gravity, or rotation velocity to the stars
\citep[see e.g.][and references therein]{Zorec17}.
    
Even if each component can be characterized by a single magnitude
and colour, its position in colour--magnitude diagrams, and hence in the
HR diagram, when translating magnitudes and colours into luminosities
and temperatures, can be altered by rotation \citep[see e.g.][]{Bastian09}. 
    
The PARSEC V2.1s tracks used during the computation of the
stellar parameters do not include rotation. For comparison
purposes we plotted the evolutionary tracks with no rotation (rotation) 
for stars with 1.7 and 2.5 $M_\odot$ and $Z\!=\!0.014$ from \citet{Ekstroem12}
as solid (dotted) blue lines in the lower HR diagram of Fig.
\ref{Fig:HR_diagrams}. This grid only includes models for stars 
with 1.5, 1.7, 2.0, and 2.5 $M_\odot$, in the range of interest 
for this work, an overly scarce sampling for our purposes, and are 
computed for rotation velocities that do not correspond to those of HR 10.
Therefore, we were not able to use them for the analysis. In any case
the deviation with respect to the PARSEC V2.1s tracks does not seem to be
dramatic.

\section{Conclusions}
\label{Sect:CONCLUSIONS}

Here we present a complete analysis of
the star HR 10 in the context of a large programme aimed at detecting
and monitoring variable metallic features superimposed on the
photospheric lines that could be attributed to exocometary events
\citep{Rebollido19}. HR 10 was singled out because of
its peculiar variability, the availability of a large amount of
high-resolution spectra in archives and publications, and the hints of
a first interferometric PIONIER/VLTI observation obtained in 2014
that suggested that HR 10 could be a binary. Dedicated campaigns, both
with that instrument and with spectrographs in several telescopes,
together with all the archival material have allowed us to carry out a
thorough study whose main results can be summarised as follows.

\begin{enumerate}
    \item Four interferometric PIONIER/VLT observations have provided
    solid evidence that HR 10 is a binary, the contrast between the
    components being  $\sim\!32$\% in $H$-band.
    
    \item The analysis of more that 32 years of high-resolution
      spectroscopic observations from archives, publications, and
      since 2015, from our dedicated campaigns, shows that narrow CS
      absorption features are present in at least 24 metallic lines
      of Ca {\sc ii}, Ti {\sc ii,} and Fe {\sc ii}.
    
    \item Particular attention has been paid in this work to the
      strong Ca {\sc ii} K CS narrow absorption features, which can be
      decomposed into two main components, labelled `D' and `S' -- for
      `deep' and `shallow'. These components show a periodic behaviour, with
      component D moving to the blue (red) as component  S  moves to the
      red (blue), crossing each other at certain times; this pattern
      being kept over decades.
    
    \item In those situations where the separation in RVs of  components D and
      S is large, two additional weaker components,
      labelled `d' and `s' are very apparent, each one to the red side
      of components D and S, respectively. These weaker d and s
      absorption features seem to accompany the main D and S in their
      periodic movements and not only appear in the Ca {\sc ii} K
      narrow absorptions, but also in lines of other species, like Ti {\sc
        ii} and Fe {\sc ii}.
    
    \item Each star holds its individual shell or envelope. The detailed
      analysis of the time evolution of  components D and S shows
      that they trace the orbit of each individual star in the
      binary. This is strongly confirmed by the spectrometric binary
      solution. The orbital period is $P_{\rm orb}\!=\!747.6$ days,
      the mass ratio $q\!=\!M_{\rm B}/M_{\rm A}\!\simeq\!0.72-0.84$,
      and the eccentricity of the orbits is $e\!\simeq\!0.23$.
    
    \item The complete orbital solution was obtained using the
      results of the RV analysis and the four PIONIER/VLTI astrometric
      points. This allowed us to compute the inclination of the
      system, $i\!=\!93.34^\circ$ and the total semimajor axis, $a_{\rm
        tot}\!=\!a_{\rm A}+a_{\rm B}\!=\!10.58$ mas, which at the
      distance to HR 10 (145.18 pc) is equivalent to $\sim\!3.08$ au;
      this implies a minimum distance between the stars -- when both
      are located at their pericentre -- of $\sim\!2.39$ au, a
      remarkably small value.
    
    \item Stellar parameters for HR 10-A and -B were estimated
      making use of observable quantities -- photometry, optical and
      ultraviolet spectra, and the constraint imposed by the PIONIER
      observations -- and synthetic low- and high-resolution spectra
      and evolutionary tracks.
    
    \item To our knowledge, this is the first case studied of a
      (slightly off)-MS binary where both components have individual
      envelopes or shells. The current analysis does not show any
      variability in the CS components that could be attributed to
      exocometary events. The absence of a narrow absorption component
      at the systemic velocity rules out the presence of a circumbinary
      envelope.
    
\end{enumerate}

As a final remark, we would like to explicitly mention the fact that 
there is a chance that other stars that exhibit variability attributed to 
FEB phenomena might actually be binaries and behave in a way qualitatively similar 
to HR 10. \citet{Marion14} estimated that when using interferometry and single-aperture 
imaging, about half the population of nearby A-type stars could be resolved as 
binaries, and suggested that a number of them remain undetected. In the case 
of HR 10, the key point that drove the whole analysis presented in this 
paper was the first PIONIER observation obtained in 2014 that 
showed strong evidence that the star was a binary. 

In a generic case, should the interferometric observations not be feasible, 
spectroscopy alone would be able to unveil the binarity, provided that 
observations obtained during a well-sampled and long time interval can be 
collected; the binarity would show up in the structure of the photospheric 
lines provided the projected rotation velocity of the stars were low enough, and/or 
in the periodic shifts of the narrow absorption component (components), 
if it (they) were present and found to show a similar shape during timescales long 
enough to rule out the FEB scenario. Otherwise, if the orbital period of the binary 
is of the order of months or years, the analysis of the variability of the narrow 
absorptions during shorter intervals (days or weeks) can be misinterpreted as 
originating from an exocometary event.

\begin{acknowledgements}
We are grateful to Prof. Barry Welsh, who refereed this paper,
for his report and comments. The authors also acknowledge 
Prof. Lynne Hillenbrand for her valuable suggestions, scientific 
discussion, and for coordinating the reduction of the HIRES archival 
spectra, and Prof. Ren\'e Oudmaijer, Dr Antonio
Garufi, and Dr Sebasti\`a Barcel\'o for their comments; we also thank 
Dr Valentin Christiaens for obtaining the 2014 PIONIER observations, 
Daniela Paz Iglesias for covering the FEROS 2015 campaign, and Dr Trifon 
Trifonov for his advice during the reduction of some spectroscopic 
data. This work is partially based on observations made with the 
Nordic Optical Telescope (NOT), operated by the Nordic Optical Telescope 
Scientific Association (NOTSA), the Italian Telescopio Nazionale 
Galileo (TNG), operated by the Fundaci\'on Galileo Galilei of the 
Istituto Nazionale di Astrofisica (INAF), and the Mercator Telescope, 
operated by the Flemmish Community; all three located at the Spanish 
Observatorio del Roque de los Muchachos of the Instituto de Astrof\'{\i}sica 
de Canarias on the island of La Palma; also on observations obtained 
with the MPIA-ESO/2.2-m telescope operated at La Silla Observatory 
by the Max-Planck-Institut f\"ur Astronomie, and observations collected 
at the European Southern Observatory with the Very Large Telescope 
Interferometer (VLTI) operated at Cerro Paranal, under ESO programmes 
093.C-0712(B), 099.C-2015(A) and 0101.C-0182(B). This paper also includes 
data taken with the 2.7-m Harlan J. Smith Telescope at McDonald Observatory 
of The University of Texas at Austin; we thank David Doss at McDonald 
Observatory for his valuable assistance in obtaining the high resolution 
spectra. Data acquired at the Anglo-Australian Telescope (AAT) have been also 
used; we acknowledge the traditional owners of the land on which the AAT
stands, the Gamilaraay people, and pay our respects to elders past and
present; we thank Stuart Ryder for his assistance at the AAT. Data
from the ESO Science Archive Facility, programmes 079.C-0789 and
094.C-0946, have been used. This research has made use of the Keck
Observatory Archive (KOA), which is operated by the W. M. Keck
Observatory and the NASA Exoplanet Science Institute (NExScI), under
contract with the National Aeronautics and Space Administration; data
from programmes U101Hr (PI G. Marcy) and C328Hr (PI J. Johnson) were
used. CE, GM, BM, IR, and EV are supported by Spanish grant AYA
2014-55840-P; HC acknowledges funding from the ESA Research Fellowship
Programme; JJEK acknowledges support from the Academy of Finland grant
295114, and thanks the Universidad Aut\'onoma de Madrid staff for
their hospitality; the research of IM is funded by a Talento Fellowship
(2016-T1/TIC-1890, Government of Comunidad Aut\'onoma de Madrid,
Spain). This work has made use of data from the European Space Agency
(ESA) missions International Ultraviolet Explorer ({\em IUE}) (archive
\url{http://sdc.cab.inta-csic.es/ines/}), and {\em Gaia}
(\url{https://www.cosmos.esa.int/gaia}), processed by the Gaia Data
Processing and Analysis Consortium (DPAC,
\url{https://www.cosmos.esa.int/web/gaia/dpac/consortium}).  This
research has made use of the SIMBAD database, operated at CDS,
Strasbourg, France.
\end{acknowledgements}

\bibliographystyle{aa} 
\bibliography{montesinosb} 

\end{document}